\documentclass{jfm}
\usepackage{graphicx}
\usepackage{epstopdf, epsfig}

\shorttitle{Growth dynamics of turbulent spots in the plane Couette flow}
\shortauthor{M. Couliou  and R. Monchaux}

\title{Growth dynamics of turbulent spots in plane Couette flow}

\author{Marie Couliou and Romain Monchaux\thanks{Email address for correspondence: monchaux@ensta.fr}}

\affiliation{IMSIA, ENSTA ParisTech, CNRS, CEA, EDF, Universit\'e Paris-Saclay, 828 Boulevard des Mar\'echaux, 91762 Palaiseau Cedex, France}

\begin{document}
\maketitle

\begin{abstract}
We experimentally and numerically investigate the temporal aspects of turbulent spots spreading in a plane Couette flow for transitional Reynolds numbers between 300 and 450. Spot growth rate, spot advection rate and large-scale flow intensity are measured as a function of time and Reynolds number. All these quantities show similar dynamics clarifying the role played by large-scale flows in the advection of the turbulent spot. The contributions of each possible growth mechanism: growth induced by large scale advection or growth by destabilization, are discussed for the different stages of the spot growth. A scenario which gathers all these elements is providing a better understanding of the growth dynamics of turbulent spots in plane Couette flow that should possibly apply to other extended shear flows.


\end{abstract}
\section{Introduction}\label{sec:intro}
Transition to turbulence in wall-bounded shear flows often occurs through subcritical scenarios when their natural control parameter, the Reynolds number ($Re$) is increased. This is the case of plane Couette flow (PCF), Hagen-Poiseuille flow, Taylor-Couette flow (when the outer cylinder is rotating) and plane Poiseuille flow. In all these cases, the laminar profile is linearly stable for $Re<Re_c$ but it is actually observed to be nonlinearly destabilized by finite amplitude perturbations above $Re_g<Re_c$. Below this threshold $Re_g$, any perturbation brought to the laminar profile eventually vanishes. Above, the flow is highly sensitive to the perturbation nature and intensity. The ultimately observed state results from the competition between coexisting turbulent and laminar areas. When $Re$ is larger than $Re_t>Re_g$, this final state consists of homogeneous featureless turbulence, while for $Re_g<Re<Re_t$, it involves both laminar and turbulent areas. This mixed state can be unpredictably evolving or steady and organized --as in the turbulent spiral formerly observed by \cite{coles65_JFM} and \cite{vanatta66_JFM} in Taylor-Couette flow and later characterized at a much larger aspect ratio by \cite{prigent02_PRL}. In both cases, the laminar-turbulent interfaces are found to be oblique, a so far unexplained feature shared by all such patterns now observed in all the extended "subcritical" flows cited above -- see \cite{prigent05_IUTAM}, \cite{barkley05_PRL} and \cite{duguet10_JFM} for plane Couette flow and \cite{hashimoto09_THMT} and \cite{tuckerman14_POF} for plane Poiseuille flow. These patterns are even observed in Waleffe flow, a shear flow without walls, as recently proven by \cite{chantry15_ARX}.

More basically, the simple development of turbulence within a formerly laminar state is far from being fully understood. From the early observations by \cite{emmons51_JAM} in boundary layers, it obviously appears that localized perturbations trigger the growth of turbulent spots that consist of simply connected turbulent regions immersed in an otherwise laminar flow and bounded by fronts that move in order to expand the turbulent area provided $Re$ is large enough. Turbulence in these kinds of transitional flows consists of long streamwise velocity streaks associated with counter-rotating vortices that are stacked together along the spanwise direction. As the wavelength of the streaky pattern is narrowly distributed around a given value $\lambda_s$, the spanwise growth of a turbulent spot corresponds to the nucleation of new streaks. The mechanisms at work to achieve this expansion have not yet been all revealed in spite of many experimental and numerical studies. Shedding light on the growth process of turbulent spots can also be a key point in understanding the formation and sustainment of organized patterns.
	\begin{figure}
	\begin{center}
	\includegraphics[width=0.8\textwidth]{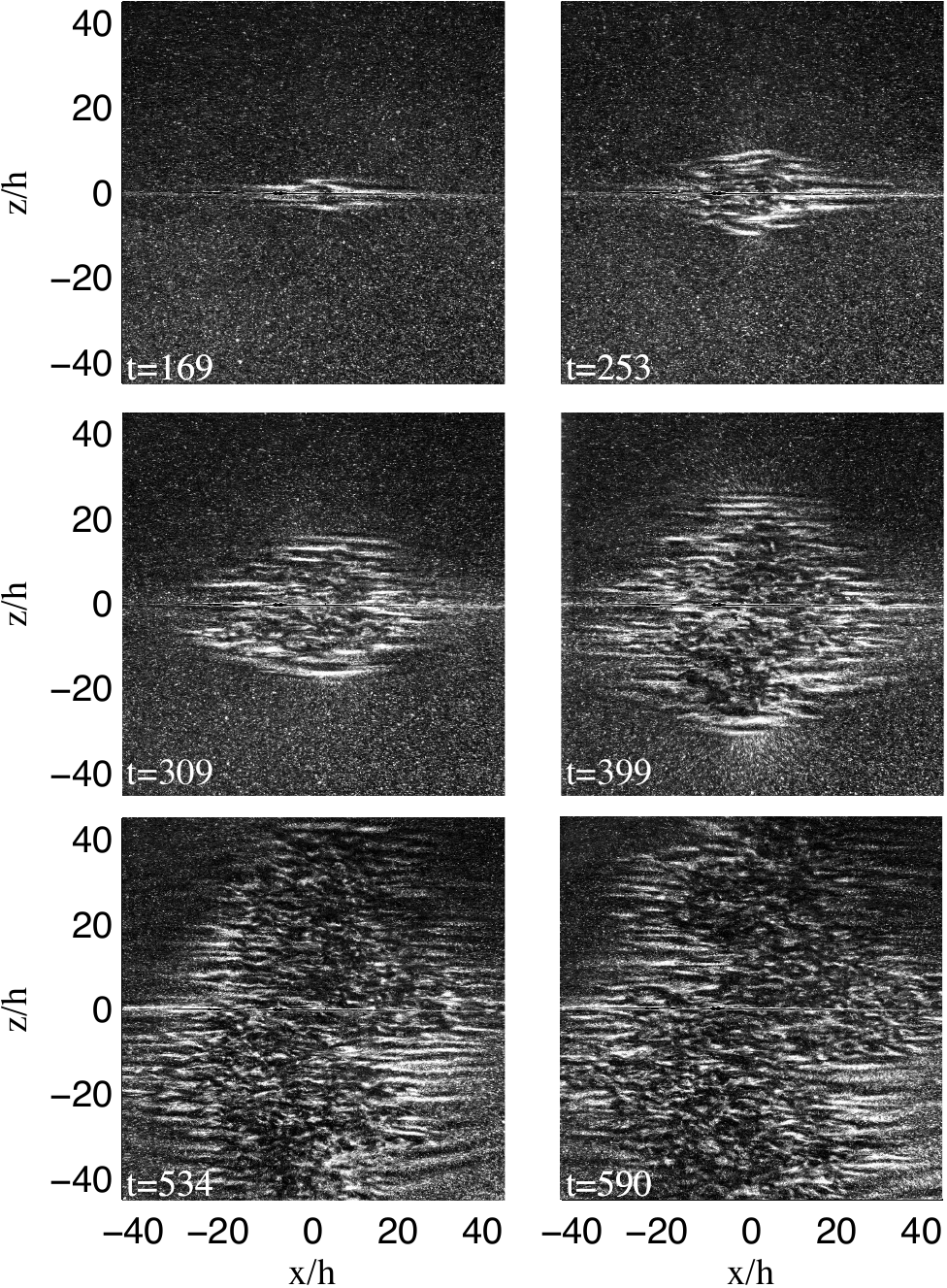}\\
	\end{center}
	\caption{Snapshots of a spot growing around a bead during a step experiment (sudden increase at $t=0$ of $Re$ from $0$ to $403$). Times are given in unit of $h/U$. First phase last till $t\simeq 400~h/U$ when the reorganization process starts. See text for details. The corresponding space-time diagram can be found in figure \ref{fig:2DST}.}
	\label{fig:Snapshots_bead}
	\end{figure}
	In the case of plane Couette flow, for which no mean advection is present, turbulent spot growth can be split into two phases: in the first stage of their development, they grow in a more or less symmetric fashion leading to diamond-like shape as seen in figure~\ref{fig:Snapshots_bead} and then, when their spatial extension is large enough, they start being distorted and tend to form oblique branches. As first evidenced numerically in the pioneering work by \cite{lundbladh91_JFM}, the spot remains centro-symmetric all along the first phase. The flow and pattern symmetries are detailed in an extensive review  \citep{barkley07_JFM}. Figure \ref{fig:Snapshots_bead} gives an illustration of the growth of such a spot in our experiment around a localized perturbation. The spot is indeed elliptic, the principal axis along the streamwise direction being longer than the one along the spanwise direction. As time passes, both principal axes tend to have the same length and the spot takes the form of a more or less regular diamond whose sides form an angle with respect to the streamwise direction. This angle is prefiguring the organized pattern orientation. This dynamic results from growth rates that are significantly larger in the spanwise than in the streamwise direction as shown successively by the two experimental works by \cite{tillmark92_JFM} and by \cite{dauchot95a_POF} confirming the numerical investigation by \cite{lundbladh91_JFM}. These researchers focused on global growth rates along both relevant directions and found in first approximation that they were time independent but increasing functions of the Reynolds number as can be seen in figure \ref{fig:cmp_kth_saclay} that gathers their time-average results of spanwise growth rate s a function of $Re$. 
	\begin{figure}
	\begin{center}
	\includegraphics[trim = 0mm 0mm 0mm 0mm, clip,width=.75\linewidth]{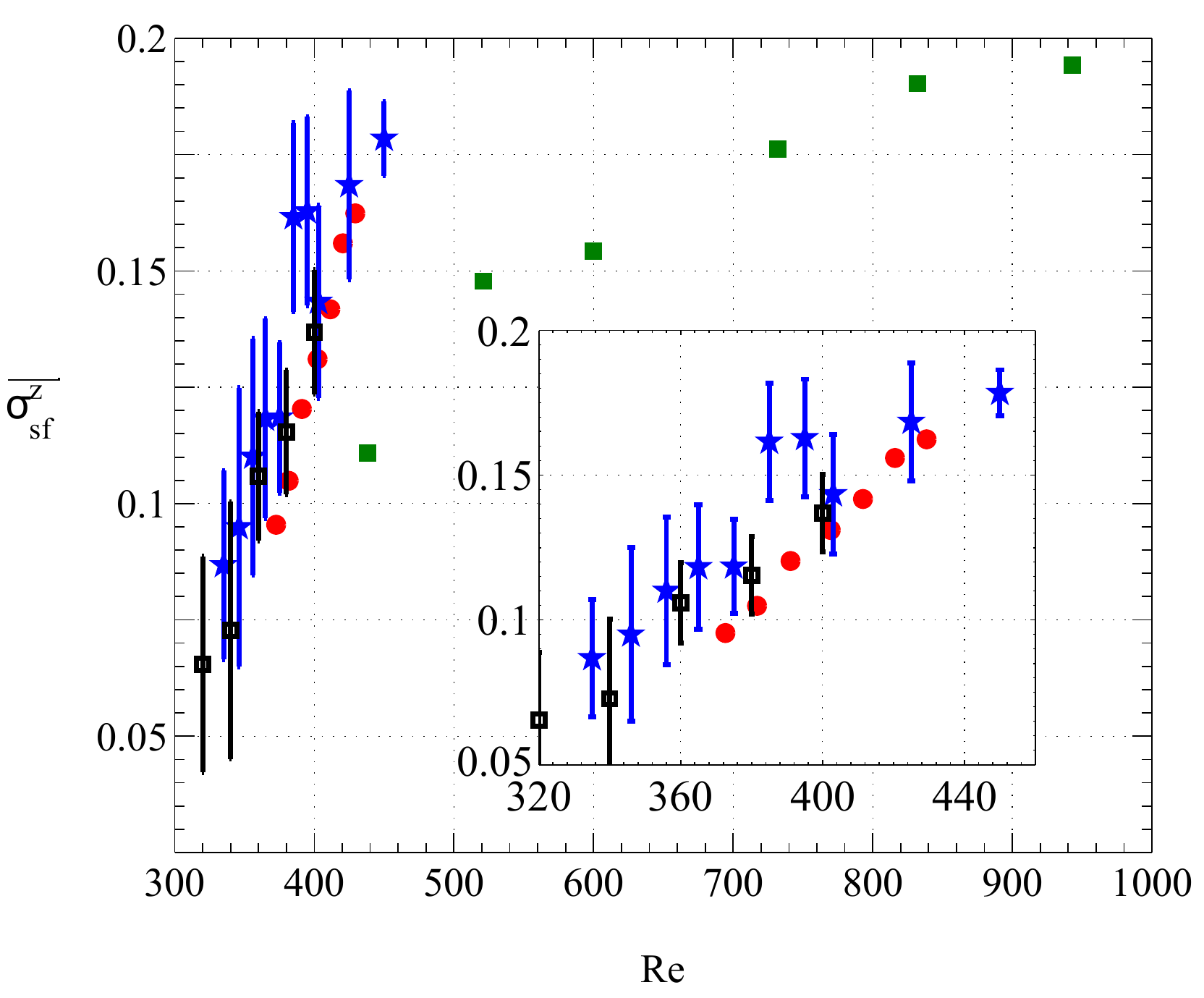}
	\end{center}
	\caption{Spot spanwise spreading rate $\overline{\sigma_{sf}^z}$ as a function of $Re$ in \cite{tillmark95_EPL} (filled squares), \cite{dauchot95a_POF} (filled circles), our experiment (filled squares with errobar) and our numerical simulation (filled stars with errobar). Note that $\overline{\sigma_{sf}^z}$ is defined in the present article as the temporal average of half the total spot spanwise growth rate, or equivalently as the average of both spanwise front velocities. Other authors' data have been obtained graphically from their publications and scaled to comply with the present definition of $\overline{\sigma_{sf}^z}$ .}
	\label{fig:cmp_kth_saclay}
	\end{figure}

	The second phase is illustrated in figure \ref{fig:Snapshots_bead} at times larger than $400$. Specific studies of this second phase are scarce. In very large domains, \cite{duguet10_JFM} nevertheless present the typical evolution of a diamond shaped spot turning progressively into a turbulent pattern made of several adjacent bands
with various orientations that are similar to the more steady patterns described above (see in particular their figures 2 and 4). In domains of moderate extension (say when the spanwise direction is $40$ times the distance between the walls), this second phase starts when the spot size becomes sizable with the domain spanwise extension and consists mostly in a reorganization of the turbulent area into an inclined pattern. In the following we will focus on the first phase.

	Understanding the mechanisms responsible for the spot growth requires taking into account all the features of the spot during this first phase. From the numerical simulations of PCF by \cite{lundbladh91_JFM} and by \cite{duguet13_PRL} in larger domains, from the model flows studied by \cite{lagha07_POF} and \cite{schumacher01_PRE} and from our own experimental study \citep{couliou15_POF}, it appears that large-scale flows are always present around the growing spot. Their origin can be explained by the laminar-turbulent coexistence and more particularly by the streamwise flow rate mismatch occurring in the overhang region where the laminar flow on one wall faces turbulent flow on the other wall \citep{coles65_JFM,lundbladh91_JFM}. Large scale flows actual shape depends on the wall normal plane of observation, but they are globally oriented in the spanwise direction and outward at the spot spanwise tips and in the streamwise direction and inward at the spot streamwise tips. Experimental works already cited also report the existence of waves at the spot spanwise tips. These waves may correspond to the advection of streamwise vortices that are actually visible on the snapshots presented in figure \ref{fig:Snapshots_bead} and that have been observed from visualizations in spanwise-wall normal plans by \cite{hegseth96_PRE}. Both these features are thought to be relevant in understanding the spot spanwise growth dynamics. Following the work of \cite{gadelhak81_JFM} on the boundary layer, the spot spanwise growth is usually attributed to local instabilities of the modified laminar profile at the spanwise laminar-turbulent interface \citep{hegseth96_PRE,dauchot95a_POF}. \cite{schumacher01_PRE} emphasize the fact that suppressing the large scale recirculation in the parallel flow of interest in their work inhibits the spot growth. The same conclusion may also hold for plane Couette flow. Trying to select the most meaningful features, \cite{duguet11_PRE} have studied the growth of turbulent spots in narrow domains for which the periodicity in the streamwise direction is achieved at a wavelength shorter than the typical spot streamwise extent. By doing so, the large-scale recirculation is suppressed and the authors found a stochastic growth resulting from the competition between streak retreat and generation events. The corresponding growth rates are found to be one order of magnitude lower than those usually measured in domains extended in both directions. It is nevertheless difficult to know if this stochastic growth is actually active in such large domains since the stochastic growth rate could be smaller than the measurement uncertainties on the total growth rate. The role of large-scale flows can primarily be seen as a disturbance to the stable laminar profile around the turbulent spot. The temporal evolution of the large-scale flows is quite slow as compared to the rapid turbulent fluctuations and happens on the same order of time scales as those associated to the spot growth as shown for example  in our experimental study \cite{couliou15_POF} or a recent model by \cite{manneville15_RNL}. As a result, the large-scale flows may induce modifications to the laminar flow around the turbulent spot that are strong enough to turn this profile unstable and steady enough to allow the instability to develop. The local corresponding mechanism is the growth by destabilization we alluded to above. A quantitative study of this mechanism has been undertaken in plane Poiseuille flow by \cite{henningson94_JEM}, but the growth rate was found too small to fully explain the experimentally observed rapid expansion of turbulent spots. If relevant, the two mechanisms we have just described are not sufficient and should be only part of the growth process. In a recent article \citep{couliou16_PRE}, we have shown the existence of two localizations for the generations of new streaks occurring during the spot growth. We have attributed these two localizations to two different mechanisms. The first one is local, happens at the spanwise tips of the growing spot and results in the nucleation of new streaks at the laminar-turbulent interface, \textit{outside} of the turbulent region. It could result from instabilities of the modified laminar flow as explained above or from the stochastic process revealed by \cite{duguet11_PRE}. The second mechanism was actually shown by our study. It occurs globally and is induced by an advection due to the large scale flows present around the spot that tend to spread it in the spanwise direction. A first consequence of this spreading is a temporary widening of the streaks constituting the turbulent spot before a new streak is nucleated by some wavelength instability \textit{inside} the turbulent phase to restore the favoured state consisting in more or less equally spaced streaks.

	The spot growth process results from the interplay between several mechanisms and involves large-scale flows. Isolating each mechanism and quantifying its contribution to the total growth rate now seems possible. In the present article, the temporal aspects of spot spreading is investigated in order to quantify to which extend the two identified growth mechanisms are present and interact to achieve the time-averaged results which we have presented in our ealier work \citep{couliou16_PRE}. To do so, we analyse experimental and numerical data which acquisition and post-processing will be described in section \ref{sec:method}. The main results consist of growth rates, front velocities and large-scale flow measurements that are gathered in section \ref{sec:res}. Discussion of these results are given in section \ref{sec:disc} before we summarize our findings and identify perspectives of this work in section \ref{sec:persp}.

\section{Methods}\label{sec:method}
	In this section, we present our experimental setup, the numerical code used to perform the direct numerical simulations (DNS), the protocols used to trigger the turbulent spots, the measurement and visualisation systems and the various post-processing techniques applied to our data.
	\subsection{Experimental setup}\label{sec:expesetup}
\begin{figure}
	\includegraphics[width=0.45\textwidth]{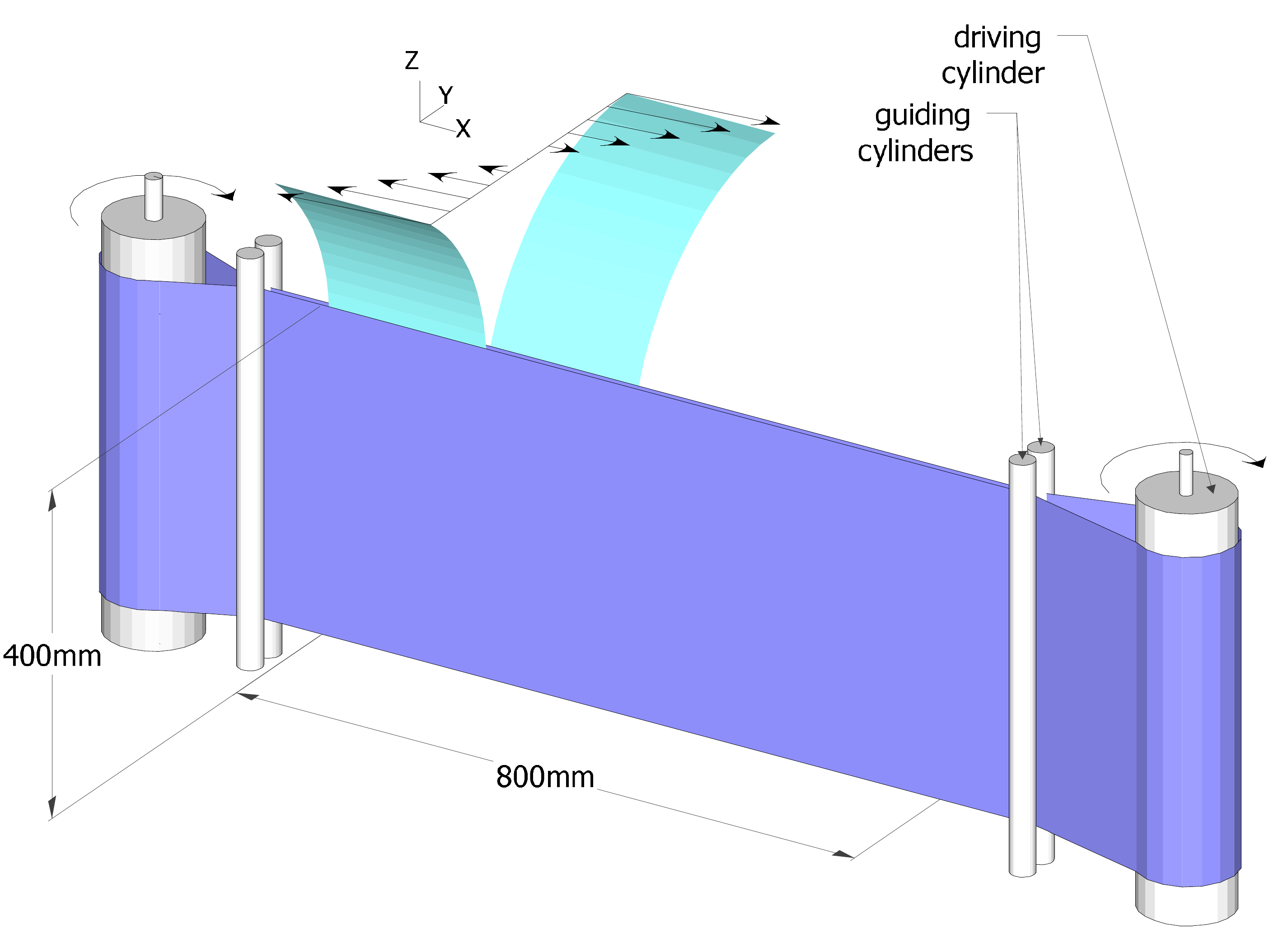}
	\includegraphics[width=0.55\textwidth]{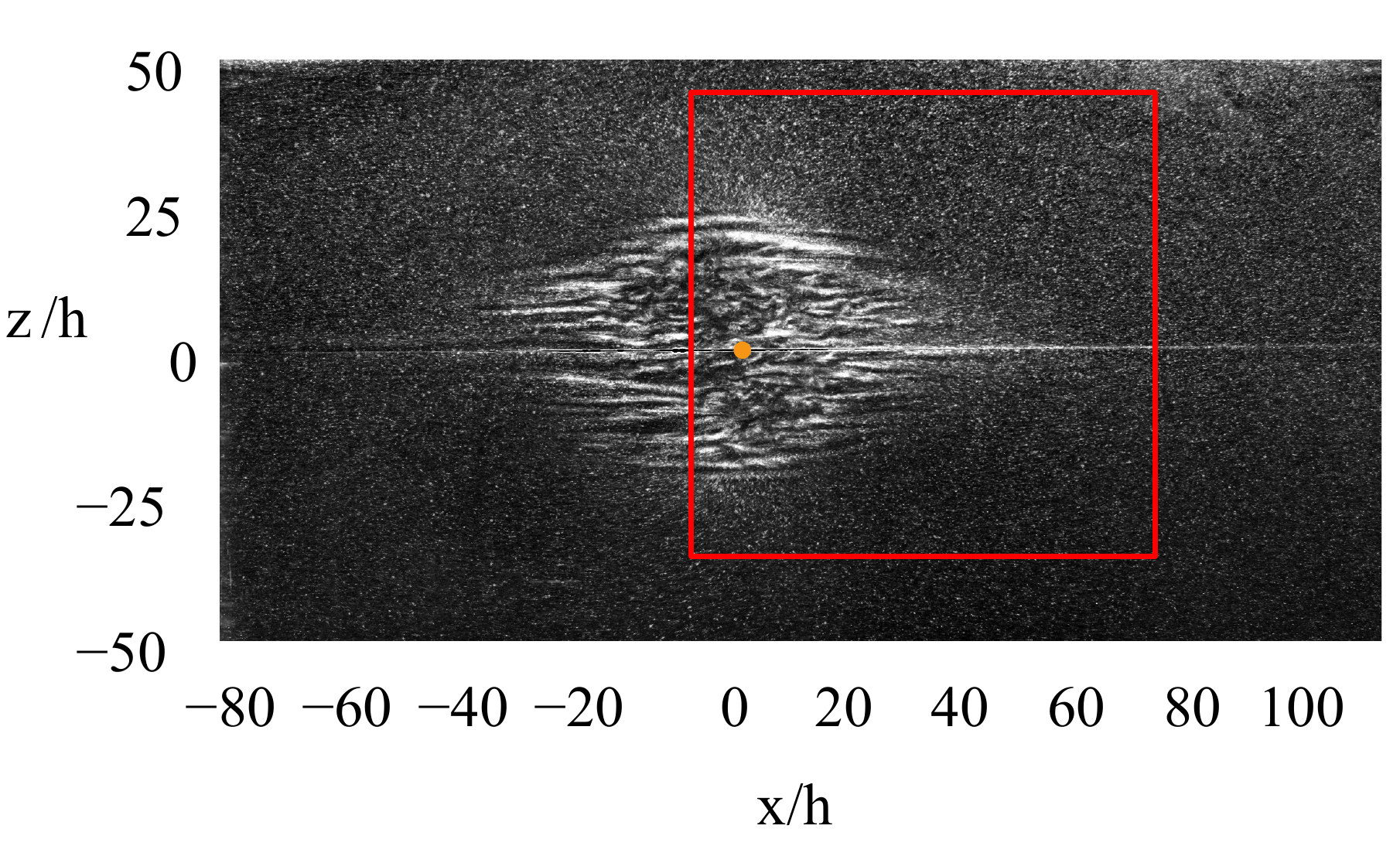}
	\caption{Left: general sketch of the plane Couette flow set-up. Right: snapshot obtained from our visualisation system during the growth of the turbulent spot. The PIV window is highlighted by the red frame. The orange dot corresponds to the position of the bead. }
	\label{fig:setup}
	\end{figure}

	As sketched in figure \ref{fig:setup}~(left), an endless $0.25$~mm thick plastic belt is stretched between two cylinders. One of them is connected to a brush-less servo-motor which drives the system through a gear-reducer of ratio $9$. This motor can achieve a maximal speed of $2500$~rpm. Speed and acceleration are controlled to an accuracy of $2/1000$. Four smaller cylinders guide the belt and enable to adjust the gap between the two walls moving in opposite directions at an accuracy of $0.05$~mm. All the results presented here corresponds to a gap of $2~h=7.5$~mm. This set-up is immersed in a tank filled with water. The water temperature is monitored by a thermo-couple with a $0.5  \ensuremath{^\circ} C$ accuracy. The Reynolds number is defined as $Re=Uh/\nu$, where $\nu$ is the kinematic viscosity of the fluid, $U$ the belt velocity and $h$ the half-gap between the two faces of the belt. The uncertainties on the belt velocity, the gap width and the temperature lead to an $Re$-uncertainty of $5$. The $x$, $y$, $z$-directions are respectively the streamwise, wall-normal and spanwise directions and $U_x$, $U_y$ and $U_z$  are  the associated velocities. Due to the experimental geometry, the plane Couette profile is achieved within an area of size $L_x*L_z$ = $800*400$~$mm^2$, which is our visualisation field of view. The corresponding aspect ratios are $\Gamma_x=L_x/h=107$ and $\Gamma_z=L_z/h=53$. The plane $y=0$ is the midway between the two plates. The tank is closed at the top by a transparent Plexiglas lid wherever the plane Couette profile is achieved in order to enable top and bottom symmetrical boundary conditions. From previous experiments not detailed here, the critical Reynolds numbers have been obtained as $R_g\simeq 305$ and $R_t \simeq 390$, which are consistent with previous experiments and well-resolved direct numerical simulations.
	\subsection{Experimental protocol}\label{sec:expe_prot}
\begin{figure}
	\begin{center}
	\includegraphics[width=0.65\textwidth]{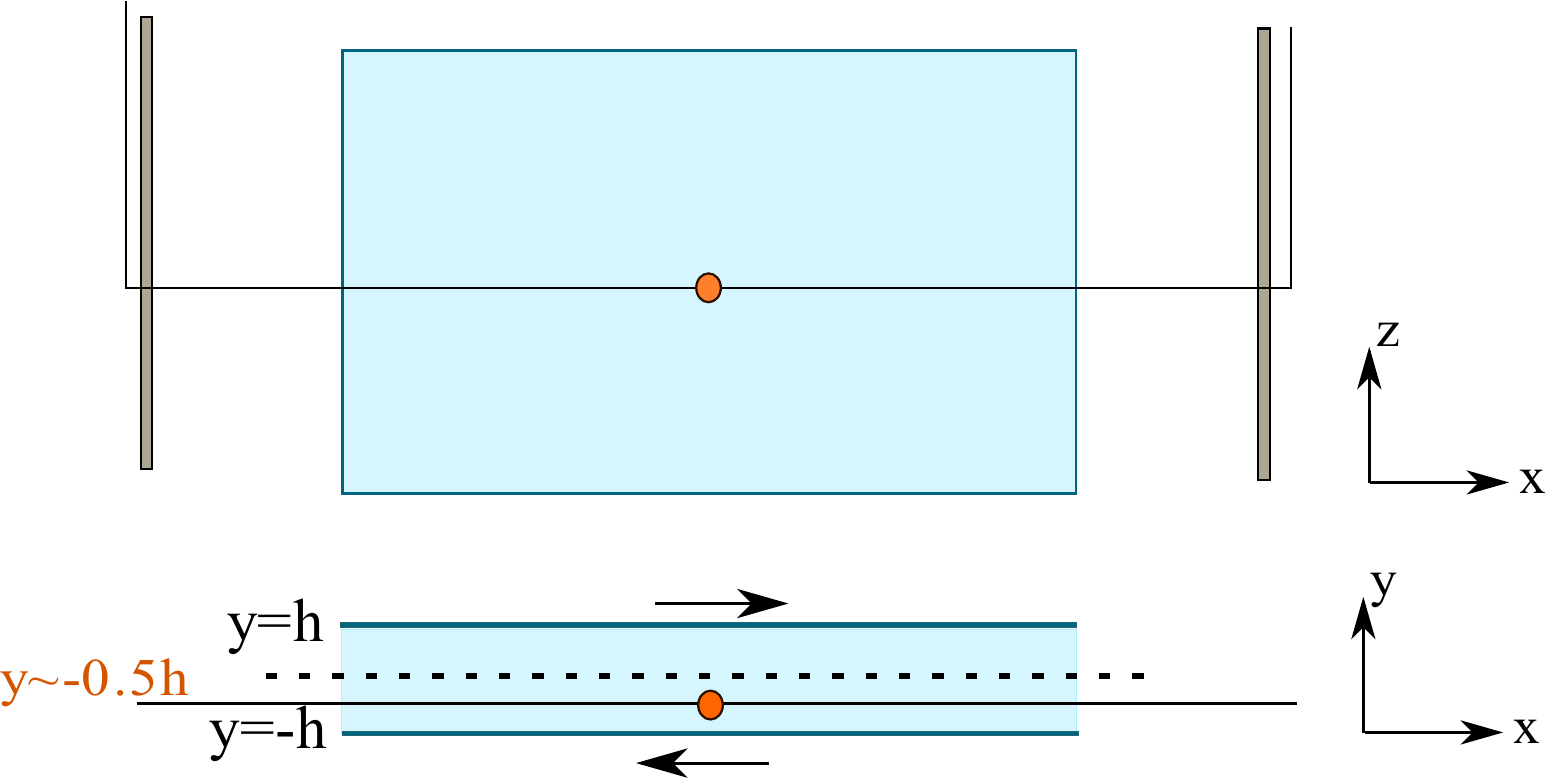}
		\end{center}
	\caption{Side and top sketch of the perturbation set-up.}
	\label{fig:bead}
	\end{figure}

	The flow is permanently perturbed by a bead positioned close to the $y=0.5~h$ plane at $(x,z)=(0,0)$ 		corresponding to the center of the experiment. The bead is maintained by means of a thin wire stretched between two stalks placed between and aligned with the guiding and main cylinders as illustrated in figure \ref{fig:bead}. This type of permanent perturbation was used by \cite{bottin97_PRL}. The wire diameter is $\phi_{wire}/h=0.1$ and the bead diameter is $\phi_{bead}/h=0.8$. As explained in \cite{bottin98_POF}, a thin wire parallel to the streamwise direction does not influence the flow since it does not induce any wake formation.The bead diameter does not strongly affect the qualitative results and provided that $Re>Re_g$, no intermittency is observed and the bead is rather only triggering turbulent spots. We studied the influence of the stalks and wire on the transition to turbulence in our system when the bead is not present: their influence is hardly noticeable when considering critical Reynolds numbers and transition time to turbulence. When turbulence develops in a formerly laminar domain, it always occurs first around the bead. A typical experiment is a step consisting in a sudden increase of $Re$ from $0$ to a final Reynolds number $Re_f$. 
	The acceleration used is such that the belt reaches its final velocity in less than $0.3$~s, \textit{i.e.} less than $10$~$h/U$, which is always very short as compared to the spot development time scales. For each kind of experiment (final $Re$ value $Re_f$, presence of the bead or not \ldots) we have performed from $5$ to $10$ independent realisations. Unless otherwise specified, results presented here are ensemble averaged over these different realisations.

	\subsection{PIV measurements and visualisations}\label{sec:PIV_Visu}
	Velocity measurements have been achieved by Particle Image Velocimetry (PIV). The Dantec system consists of a dual pulse Laser (Nd:YAG, $2\times135$~mJ, $4$ ns, $532$~nm) and a CCD camera (FlowSenseEO, $4$ Mpx). Image pairs are acquired at a rate of $5~$Hz. The PIV Laser is mounted above the test section on a linear traverse such that velocities over different $y$ planes can be measured with a spatial resolution of about $1.3$~mm along $y$. The laser sheet produced from a cylindrical lens gives $2$D velocity fields ($U_x$,$U_z$) in the streamwise/spanwise planes. The flow is seeded with particles coated with Rhodamine, the diameter of which $d_p$ is in the range $=1-20~\mu m$ and that diffuse light at a shifted wavelength. A filter mounted on the camera lens is used to keep only wavelengths around this shifted wavelength, getting rid of unwanted light reflections on the vessel and plastic belt. The $1948\times 2048$ pixels$^2$ observation window corresponds to physical sizes of $305\times 310$~mm$^2$. An adaptive cross-correlation processing is applied to an initial interrogation area of $64\times 64$ pixels$^2$  followed by a  final interrogation area of $32\times 32$ pixels$^2$ with $50\%$ overlap. The corresponding spatial resolution is $2.45$~mm, {\it i.e.} $0.6$~$h$. Each instantaneous velocity field is further filtered by removing values exceeding $1.2$~$U$. Resulting outliers are replaced using local median filters.
	
	Alternatively, visualisations are performed over the entire area where a plane Couette flow is achieved. To that purpose, the flow is seeded with Iriodin particles and lighted with a Laser diode combined with a polygon mirror rotating at high frequency to produce a laser sheet which illuminates the flow along a plane at a given $y$. Iriodin particles are micron-sized platelets which tend to align with the flow stream. These platelets respond to local flow changes and reveal laminar and turbulent regions: the turbulent state is indicated by quick fluctuations in the reflected light intensity while the laminar state corresponds to more uniform light reflections. Images at a resolution of about $2200\times 1100$ pixels$^2$, corresponding to an observation window of $800\times 400$~mm$^2$ (exactly fitting the area where the laminar Couette profile is achieved), are captured by a PCO sCMOS Camera provided by PhotonLine. Figure \ref{fig:setup}(right) illustrates typical full size visualisations with both laminar and turbulent phases coexisting.
	
Growth of turbulent spots in the whole flow are obtained from visualisations and will be presented in \ref{sec:spread:exp}.	PIV measurements are used to study large-scale flows; associated results will be presented in section \ref{sec:LSF}.  For practical reasons, it is not possible to perform PIV measurements and visualisations at the same time. As a consequence, from the experimental data, it will not be possible to match the time evolution of large-scale flows and of spot growth. We will thus mostly rely on the numerical data for such comparisons.
	\subsection{Direct numerical simulations}\label{sec:DNS}
	DNS of a plane Couette flow are computed using the Channelflow software written by John F. Gibson \citep{channelflow,gibson08_JFM}. The software solves Navier-Stockes equations numerically by using pseudo-spectral methods for the spatial discretisation with a Fourier decomposition in the ($x$,$z$) directions and Chebyshev polynomials in $y$ direction. The boundary conditions are periodic in the ($x$,$z$) directions and no-slip conditions are imposed  at the walls $i.e$ at $y=\pm1$. Numerical simulations are performed in a domain of size ($L_x$,$L_y$,$L_z$)=(180,2,80) $h$. In order to resolve all the relevant modes of turbulent flow in the $Re$ range studied, the numerical resolution is (768,33,384) dealiased modes in the ($x$,$y$,$z$) directions. According to \cite{philip11_PRE} this is sufficient to observe pattern formation. A time-step of $0.01$ was used resulting in a CFL number less than $0.6$. A pertubation consisting of four pairs of counter-rotating vortices like that used by \cite{lundbladh91_JFM} is introduced as an initial disturbance to trigger turbulent spots. As for the experiment, five realisations of the DNS have been performed for each $Re$ value and ensemble-averaged results are presented. Note that for the DNS, a slight change in the perturbation amplitude (typically $0.1\%$) is introduced to achieve some variability between realisations.
	\subsection{Space time Diagrams and front detection}\label{sec:DST}
\begin{figure}
	\begin{center}
	\includegraphics[trim = 4mm 0mm 0mm 4mm, clip,width=.495\linewidth]{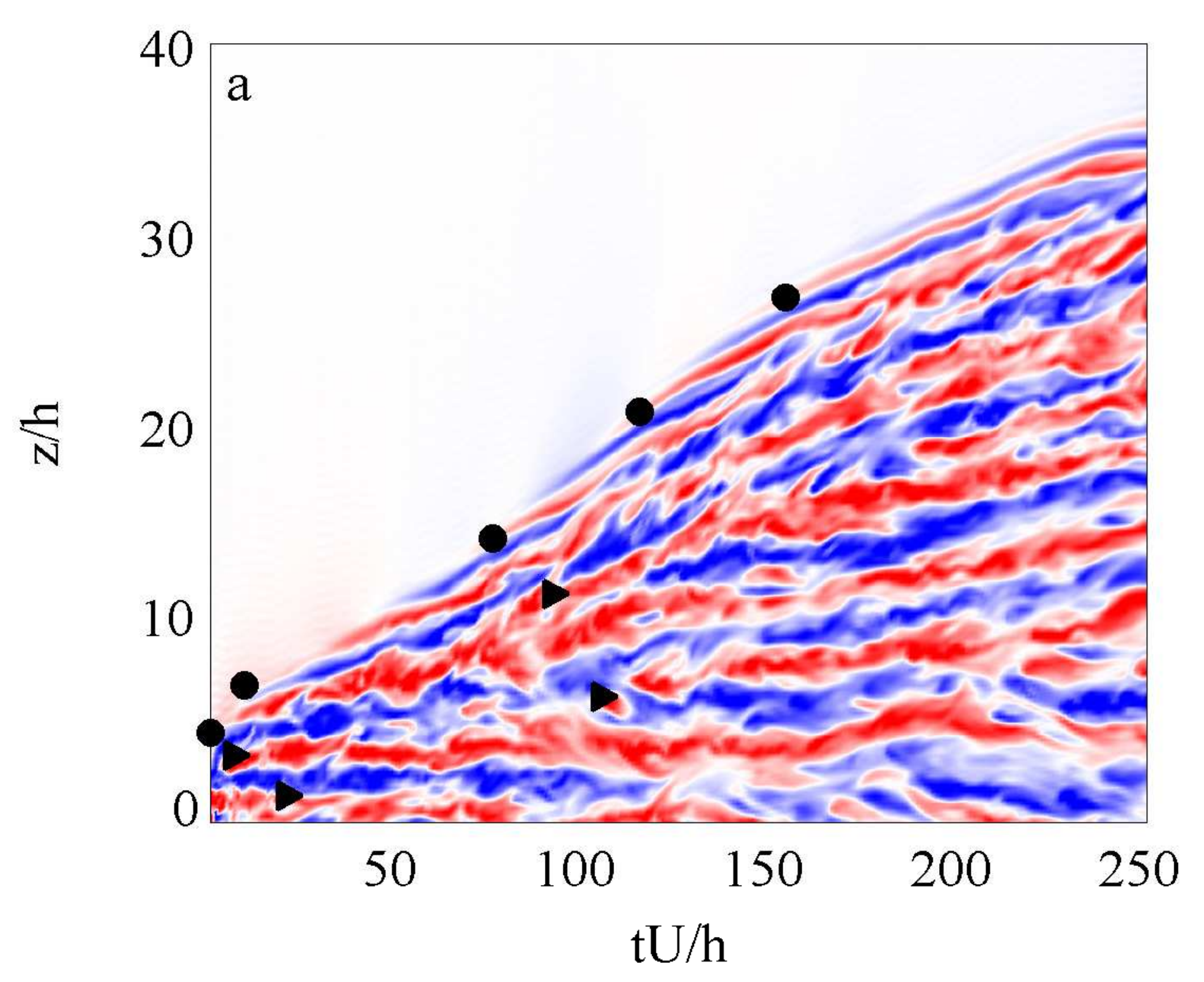}
	\includegraphics[trim = 4mm 0mm 0mm 0mm, clip,width=.48\linewidth]{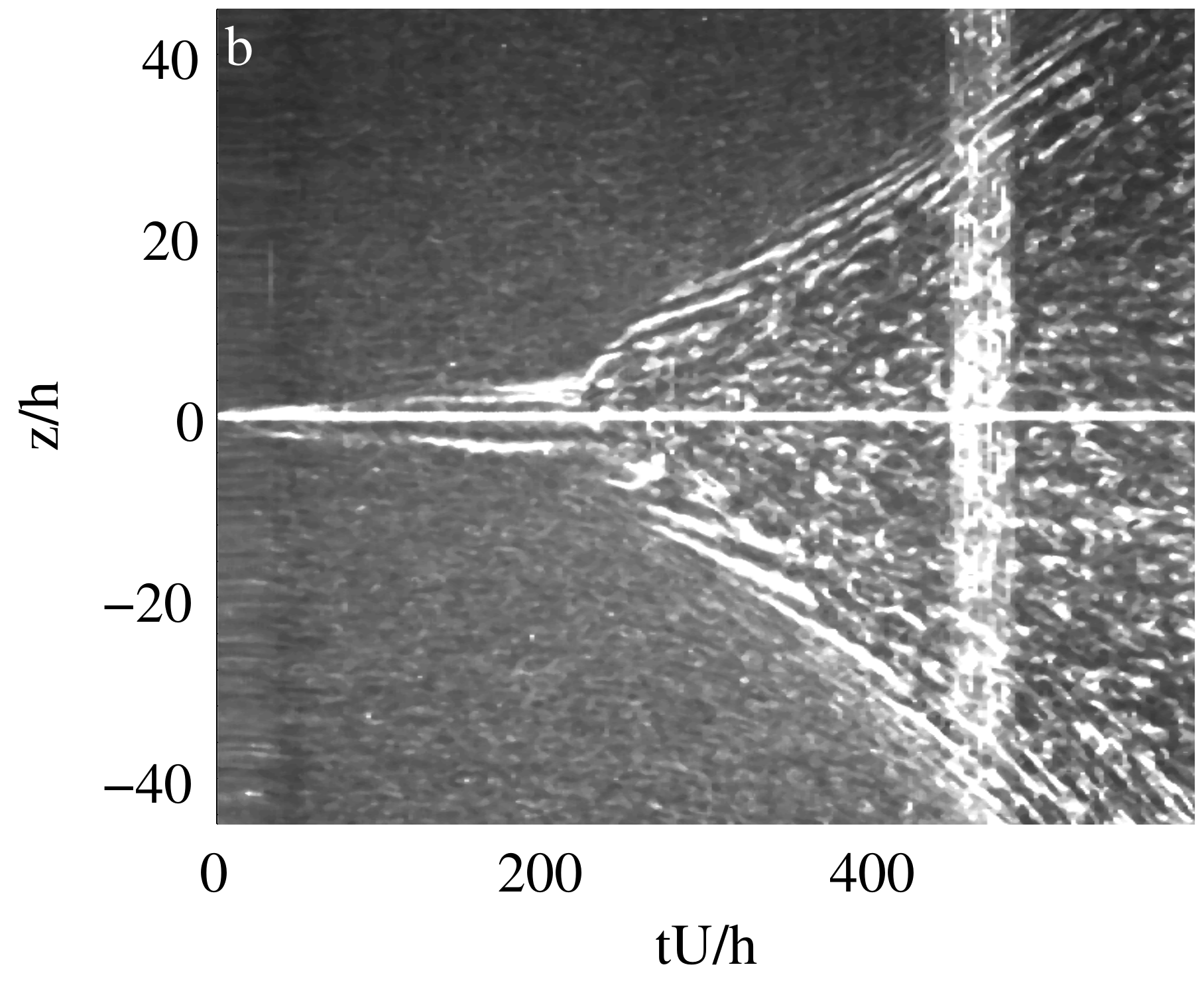}
	\end{center}
	\caption{Left: $U_x/U$ space-time diagram for a DNS at $Re=380$. Nucleations of streak inside ($	\triangleright$) and outside ($\circ$) of the spot are labelled. Only half the spot is presented in the spanwise direction. Blue: negative velocities, red: positive velocities, white: zero velocity. Right: Space-time diagram from the experimental visualisation corresponding to the growing spot presented in figure \ref{fig:Snapshots_bead}. Final $Re=403$ is reached after $10~h/U$ and, according to \cite{tillmark92_JFM}, the linear Couette profile is established after about $0.25~ Re h/U$ corresponding to $100~h/U$ in this case. The white vertical band from 450 $h/U$ to 480 $h/U$ corresponds to the tape which closes the plastic belt.\label{fig:2DST}}
	\end{figure}

	A large part of the results presented in the next section are obtained from space-time diagrams drawn from both numerical and experimental data. They are obtained from the dimensionless streamwise velocity component $U_x/U$ extracted on the line $(x,y)=(0,0)$ in the numerical case and from the $x=0$ line of light intensity acquired by our visualisation system in the experimental case. To strengthen the detected structures and in order to improve the signal-to-noise ratio, sliding averages are applied along the $x$ coordinates over $3$ samples and along the time coordinate over $5$ samples. Typical examples of numerical and experimental space-time diagrams are provided in figure \ref{fig:2DST}.

	Laminar-turbulent fronts are manually tracked in the experiment and automatically detected from the numerical data by defining two possible states: the laminar one which corresponds to the area where $U_x=0$ and the turbulent state which corresponds to the remaining area. We focus on the spanwise expansion on line $x=0$ and one can note that at this location the large-scale flow is mainly oriented toward $z$ and is consequently not that visible from $U_x$. 
	\subsection{Large-scale flows}\label{sec:meth:lsf}
	As reported by many authors (see section \ref{sec:intro}), large-scale flows develop around growing turbulent spots in shear flows. They are easily identified by studying the two-dimensional spatial power spectrum of either velocity component in the $(x,z)$ plane. These power spectra show an obvious scale separation between two peaks; one occurs around the wavelength $\lambda\simeq 4-5$~$h$ corresponding to turbulent streaks and another one, associated with large-scale flows, is situated around $\lambda=40$~$h$ \citep{duguet13_PRL,tuckerman08_JPCS,couliou15_POF}. In this work, we will use two quantities associated with large-scale flows: $A_z^{LSF}=kE_z(\lambda=40~h)/U^2$ that represents the $U_z$ premultiplied power spectra amplitude at $\lambda=40$~$h$ and $M_z^{LSF}=max(\widehat{U_z}(x=0,y=0,z),z)$, the maximal amplitude of $\widehat{U_z}$ which is the spanwise velocity component filtered above the cutoff wavelength $\lambda_{cut}=24$~$h$ on the line $x=0,y=0$. More details can be found in the work by \cite{couliou15_POF}.
	%

\section{Results}\label{sec:res}
	\subsection{General features}\label{sec:gen_feat}
		\subsubsection{Growth time scales and region definition}
		Due to our numerical and experimental protocols, the growth of the triggered spot differs slightly in the two cases presented in figure \ref{fig:2DST}. In the numerics, a turbulent zone is established very quickly (few $h/U$) from the initial perturbation that is already large, of the order of $5h$. The spot growth in our experiment takes place in two phases as can be seen in the space-time diagram of figure \ref{fig:2DST}~(right). During a first phase lasting $t_p$~$h/U$ ($t_p$ being about $210$ in the displayed example), structures develop around the bead. Then an actual spot, with an interior made of turbulent fluctuations like the ones displayed in any numerical or experimental study of such spots, grows until the end of the acquisition. In the experimental case, as our step protocol starts at $Re=0$, a time $t_c$ is required for the laminar profile to establish itself. An important point is to know how the time $t_c$ compares to $t_p$ associated with the development of the perturbation before turbulence is triggered. According to measurements performed by Tillmark and reported in \cite{tillmark92_JFM}, the linear Couette velocity profile establishes in a time $t_c\simeq 0.25$~$Re$~$h/U$. In our experiment, this corresponds to a time of the order of $100$~$h/U$ which is short enough as compared to $210$~$h/U$ when the spot is triggered.
		
	The experimental space-time diagram presented in figure \ref{fig:2DST}(right) allows us to define three distinct zones during the growth of a turbulent spot. The region connected to the bead is turbulent and displays fluctuating disorganized structures. Far away from the bead, the flow is laminar which results in roughly homogeneous dark areas. Between these two regions, long coherent inclined structures are visible. They correspond to the waves evoked by many authors as explained in the introduction (see section \ref{sec:intro}) and it has been argued that these waves correspond to the advection of streamwise vortices by the large-scale flows present at the spot spanwise tips \citep{couliou16_PRE}. A similar distinction is actually also observed on the numerical space-time diagram.

		\subsubsection{From fronts to spreading rate}\label{sec:res:front}
		
							\begin{figure} 
	 \begin{center}
	 \includegraphics[clip,trim = 4mm 0mm 0mm 30mm,width=0.495\linewidth]{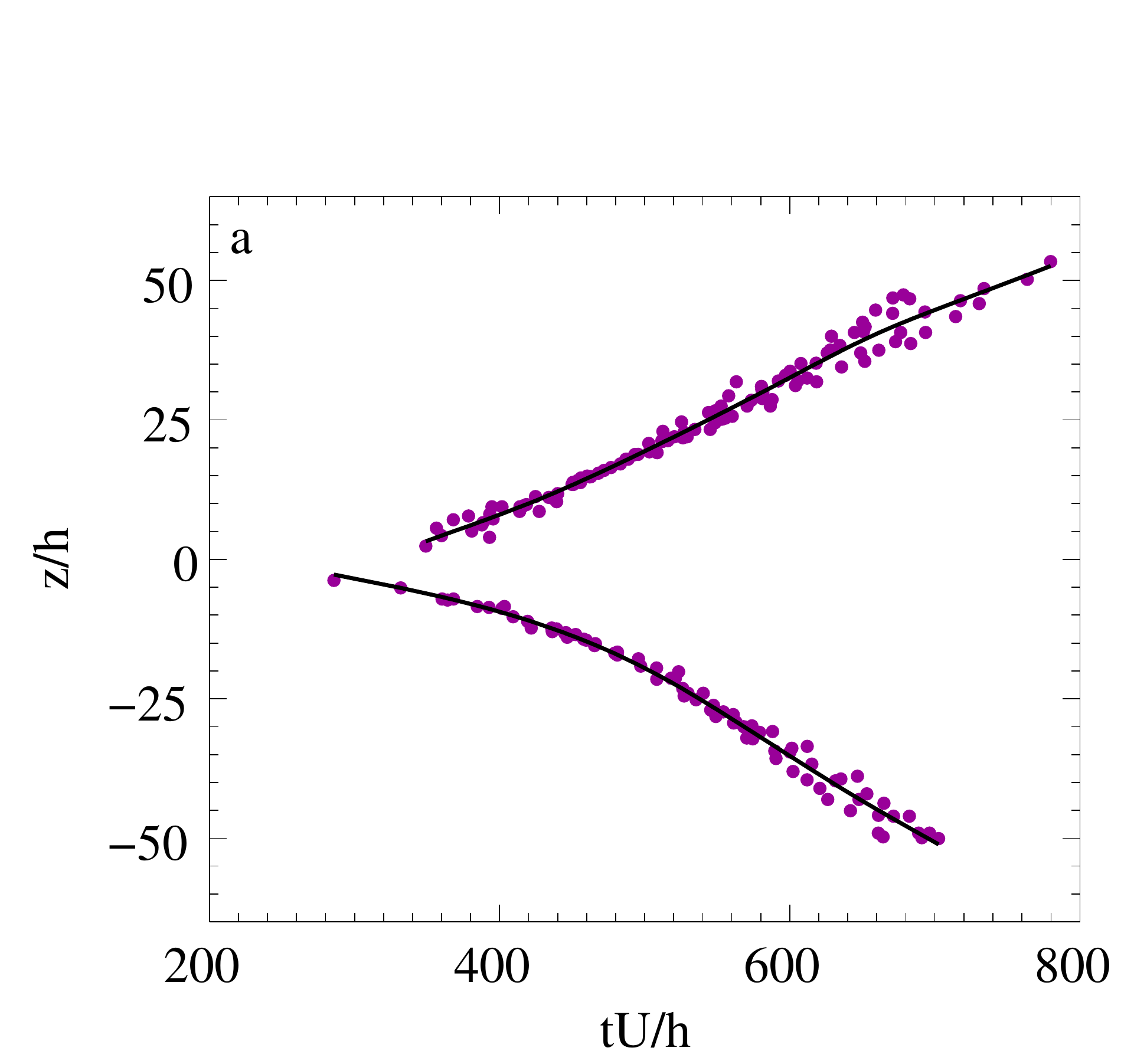}
	 \includegraphics[clip,trim = 4mm 0mm 0mm 30mm,width=0.49\linewidth]{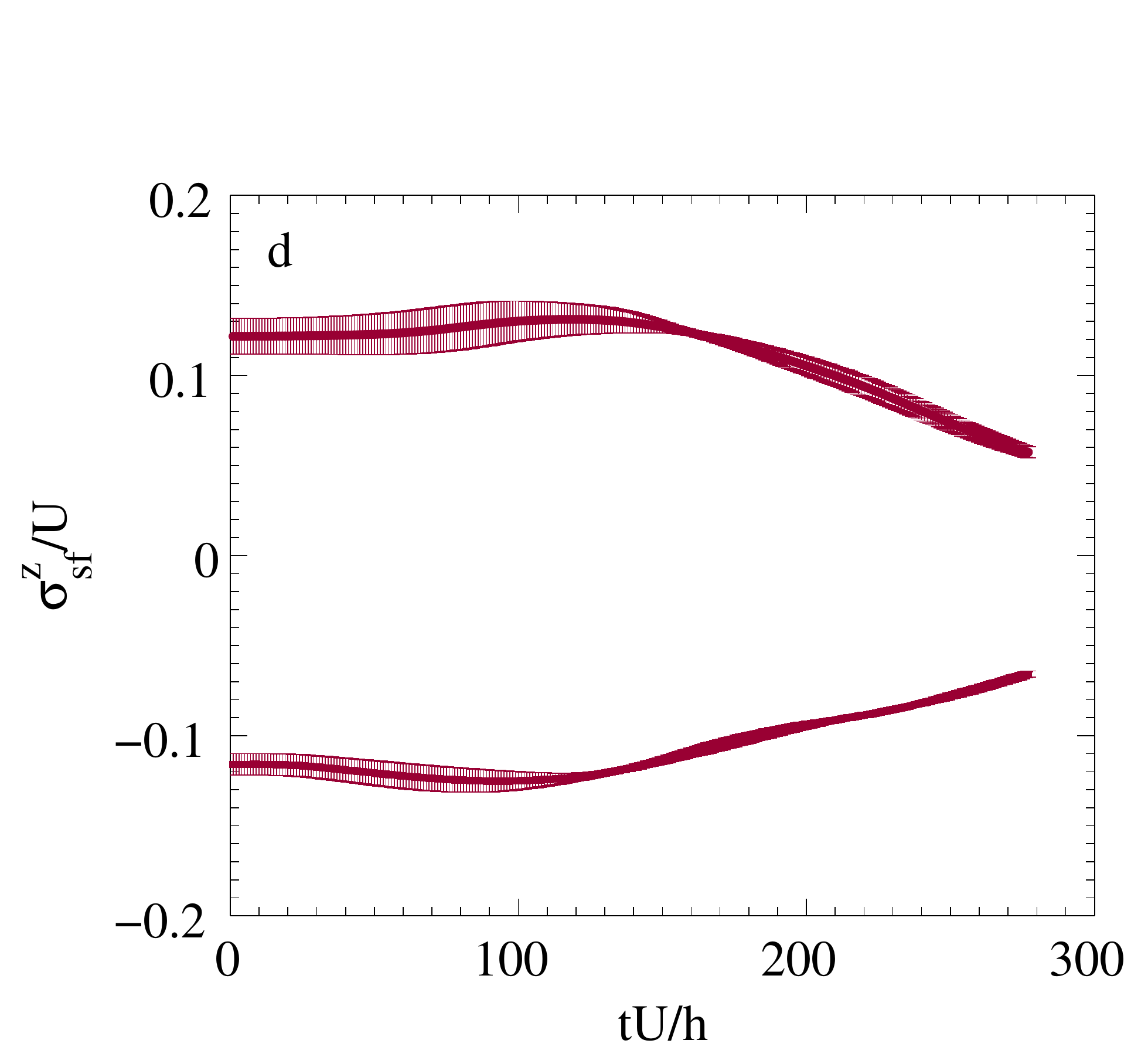}
	 \includegraphics[clip,trim = 4mm 0mm 0mm 30mm,width=0.48\linewidth]{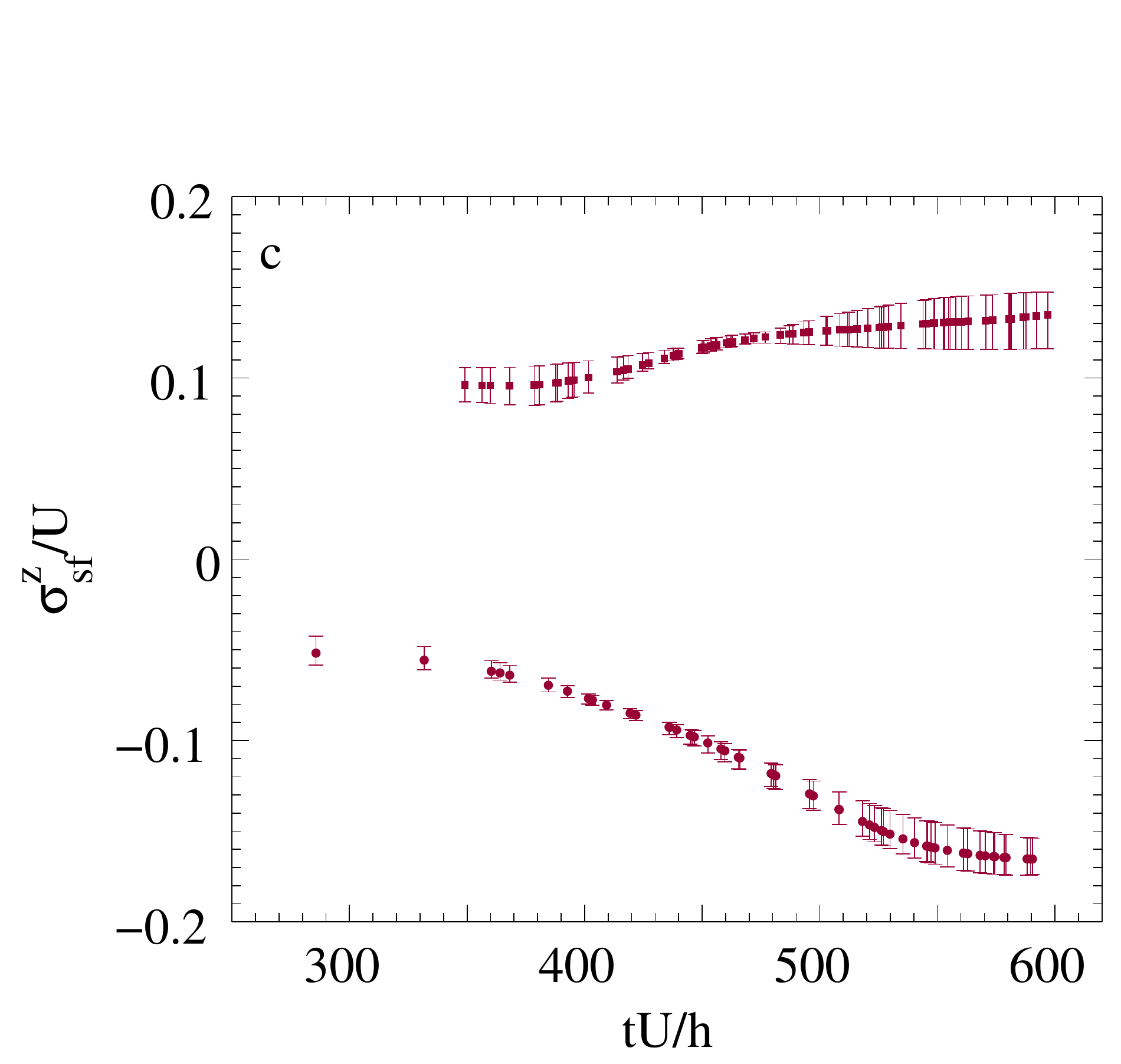}
	 \includegraphics[clip,trim = 4mm 0mm 0mm 30mm,width=0.495\linewidth]{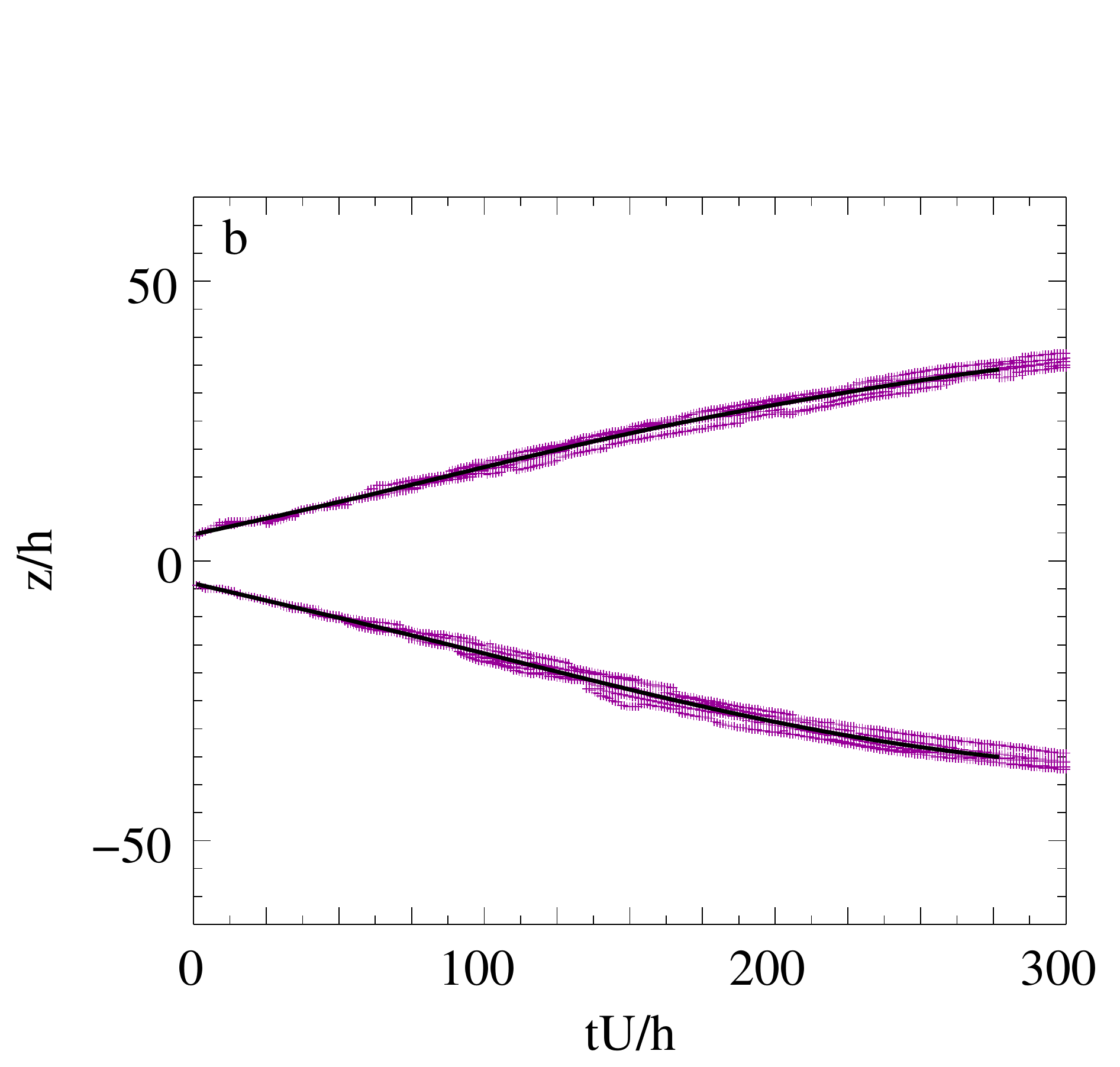}
	 \caption{Methodology figures: (a) front position as dots and corresponding smoothing spline as black line from an experiment at $Re=365$. (b) same in a numerical simulation at $Re=380$. (c) Front velocities obtained by derivating the cubic spline of figure (a). (d) Front velocities obtained by differentiating the cubic spline of figure (b).\label{fig:res:scatt}}
 	\end{center}
	\end{figure}
		The spanwise fronts of the turbulent spot are tracked to measure the spot front velocity. From the individual fronts detected manually from the experiment and automatically from the DNS as explained in section \ref{sec:DST} we obtain scatter plots such as those presented in figure \ref{fig:res:scatt}(a) and \ref{fig:res:scatt}(b) that provide the front position as a function of time for each experiment or simulation. In the experimental case, the time at which the spot starts growing may vary from one realization to another. To obtain the scatter plot presented in figure \ref{fig:res:scatt}(a), we have translated each front so that they coincide when they have reached $z=\pm 16$~$h$. Note that when $Re<320$, it may happen that no spot grows in some cases. The results presented in the following are obtained by gathering fronts associated with spots that eventually grow. In both numerical and experimental scatter plots, gathering from $5$ to $10$ independent realizations, the dispersion between different spot fronts is small enough such that an average front can be obviously distinguished. The position of this average front is obtained from cubic spline interpolation with a smoothing parameter $\mu$. This cubic spline is further differentiated to obtain the average front velocity, \textit{i.e.} half the average spot spreading rate, as a function of time at a given $Re$. In practice, we perform this velocity calculation by regularly varying the smoothing parameter $\mu$ from $0.5$ to $0.9$. We thus obtain several time velocity signals that slightly differ from one another. The ensemble average of these velocity signals is used as the spot front velocity and their ensemble standard deviation as the corresponding error bar. Figure \ref{fig:res:scatt}(c) presents such a velocity signal and its error bar for $Re=365$. We can also appreciate on this figure the relative symmetry of both fronts (top and bottom of the experiment) that is representative of the symmetry obtained at any $Re$. Note that the numerical front velocity signals are not found to be more symmetric than the experimental ones, even though the numerical boundary conditions are perfectly symmetric. Consequences are twofold: the top-bottom symmetry is fairly preserved in our experimental setup and the remaining asymmetry in the growth dynamics has to be attributed to the dynamics itself and not to the boundary conditions.

		\subsubsection{ Velocities of edge vortices}\label{sec:res:vort}
		From experimental space-time diagrams as the one presented in figure \ref{fig:2DST}(b), the edge vortices are manually detected as straight lines. Their individual velocity is then estimated as the corresponding slope. For a given $Re$, all detected velocities of the vorticies are gathered in a scatter plot as done in figure \ref{fig:res:scatt} for spot fronts. It appears that, as time passes, the velocity increases linearly and a linear fit of the scatter plot is thus performed.

		\subsubsection{Quantities of interest and relation to growth mechanisms}
		To study the spreading dynamics, we consider four quantities: $\sigma _{sf}^z$, the velocity of the spot front, $\sigma_{adv}^z$, the velocity of the vortices, $\sigma_{loc}^z$ defined as $\sigma_{sf}^z-\sigma_{adv}^z$ and the ratio $\sigma_{loc}^z/\sigma_{sf}^z$. All these quantities depend on time and $Re$. Their time averages have been studied by \cite{couliou15_POF} to reveal the two growth mechanisms described in the introduction. $\sigma _{sf}^z$ corresponds to the spot spreading rate in the spanwise direction. $\sigma _{adv}^z$ is associated with the global growth induced by large-scale flows and corresponds to the rate of streak nucleations inside the turbulent spot. From its definition, if only these two mechanisms are at work, $\sigma _{loc}^z$ measures the importance of the local mechanism and corresponds to the rate of streak nucleations outside the turbulent region. $\sigma_{loc}^z/\sigma_{sf}^z$ simply quantifies the relative importance of both mechanisms and can be compared to the fraction of nucleations occurring outside or inside the turbulent spot, which can be obtained from the numerical data. Next we present the time dependence of these four quantities in our experiments and in our simulations. In the sequel, $\overline{X}$ will denote the time average of the quantity $X$. The different quantities defined above are recalled in table \ref{tab:sigma}.\\
\begin{table}
\begin{tabular}{ccc}
	Quantity		& Definition	 &  Measurement\\ 
	\hline 
	$\sigma_{sf}$   & Total growth rate of the spot & From laminar-turbulent front velocity \\ 
	\hline 
	$\sigma_{adv}$  & Growth rate fraction due to advection & From the velocity of edge vortices\\ 
	\hline 
	$\sigma_{loc}$  & Remaining growth rate fraction  & as $\sigma_{sf}-\sigma_{adv}$ \\ 
	\hline 
\end{tabular} 
\caption{Table gathering the different growth rates used in the text, their definition and the way they are measured.}\label{tab:sigma}
\end{table}

		\subsubsection{Influence of box size and reproducibility}
\begin{figure} 
	 \begin{center}
	 \includegraphics[clip,trim = 4mm 0mm 4mm 4mm,width=0.495\linewidth]{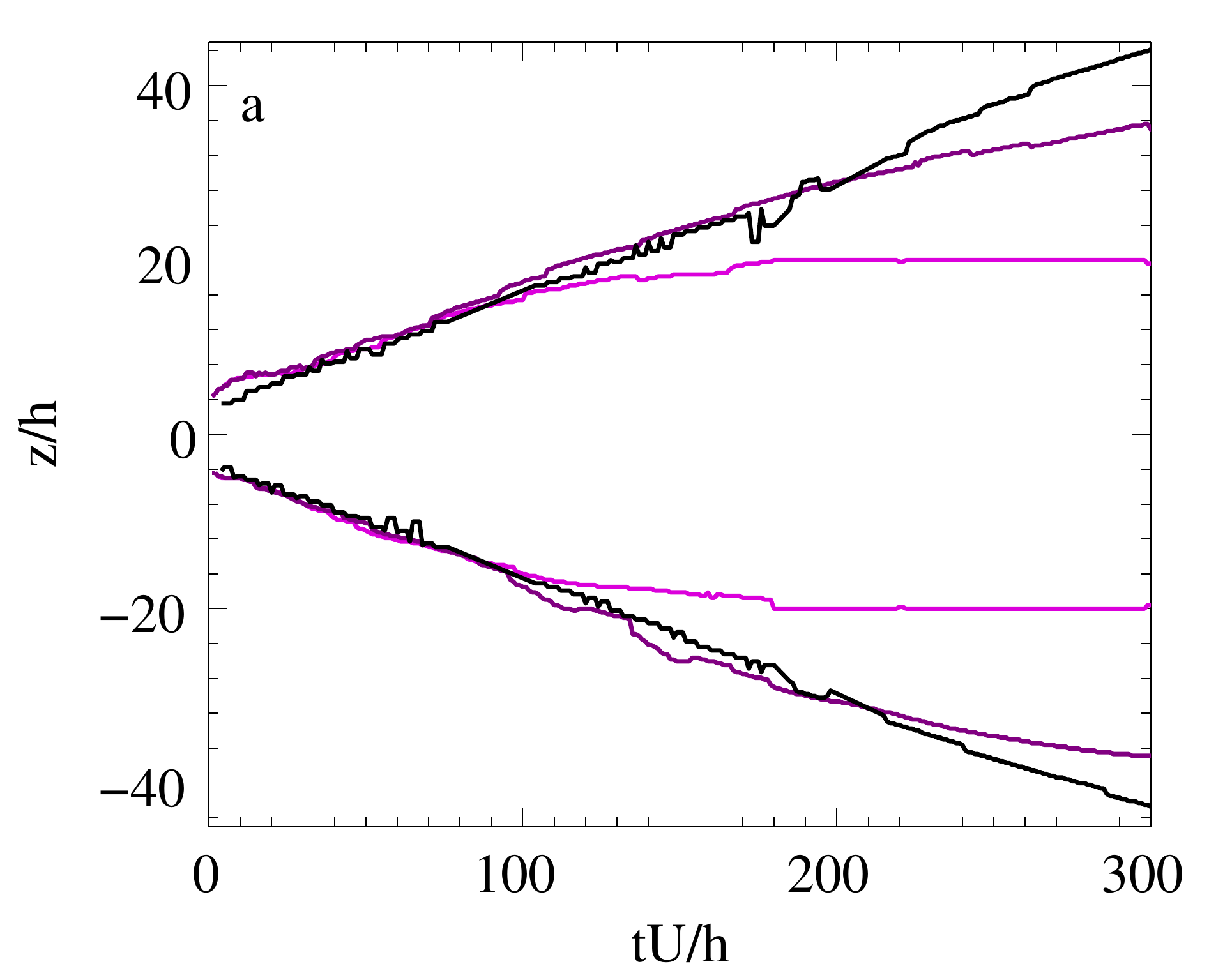}
	 \includegraphics[clip,trim = 4mm 0mm 4mm 4mm,width=0.495\linewidth]{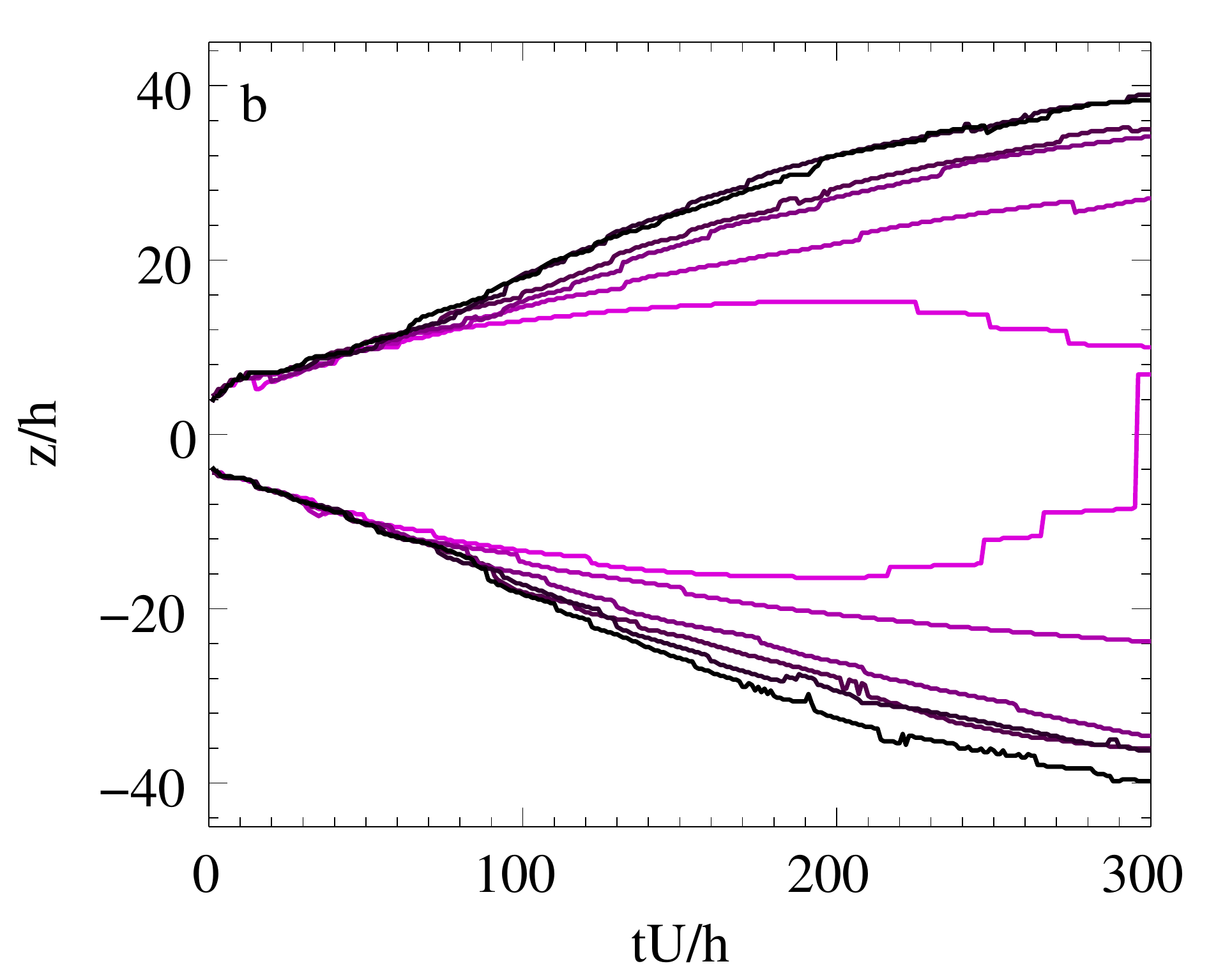}
	 \caption{Methodology figures: (a) Front position for three numerical simulations performed at $Re=380$ for three different domain size ($L_x*L_z=360~h*160~h$, $L_x*L_z=180~h*80~h$ and $L_x*L_z=90~h*40~h$). Lightest lines correspond to smaller boxes. (b) Superposition of all fronts obtained at various $Re$ by numerical simulations for a domain size of $L_x*L_z=180~h*80~h$ . From brightest to darkest, $Re$ increases from $320$ to $420$ by a step of $20$.\label{fig:res:scatt2}}
 	\end{center}
	\end{figure}

		Figure \ref{fig:res:scatt2}(a) illustrates the spot fronts for three numerical simulations performed at the same Reynolds number $Re=380$ but with increasing box sizes at constant aspect ratio. The darkest lines correspond to the spot fronts for the largest boxes. The brightest lines correspond to a smaller box than that used here and the darker to a larger box. It appears that the influence of the box size is negligible at the beginning of the growth, but becomes significant at later times when effect of the periodic boundary conditions on the spot front is not negligible any more. In order to avoid any bias from this size effect, we have decided to study the spot growth dynamics before the box size influences the spot development. More specifically, the recording of the front position should be stopped when it reaches $\pm 36h$ with our box size. The same restriction has been applied to our experimental data.

		Figure \ref{fig:res:scatt2}(b) illustrates the average spot fronts for the six Reynolds numbers studied numerically. It appears that one fourth of the dynamics are identical, regardless of the Reynolds number and even when the spot eventually vanishes (see the brightest lines corresponding to $Re=320$). During this first phase, the spot growth rate is roughly independent of $Re$ and so are the mechanisms responsible for the spot spreading that can possibly involve non-normal growth.
	\subsection{Spreading in experiments}\label{sec:spread:exp}
\begin{figure} 
	 \begin{center}
	 \includegraphics[clip,trim = 4mm 0mm 0mm 30mm,width=0.495\linewidth]{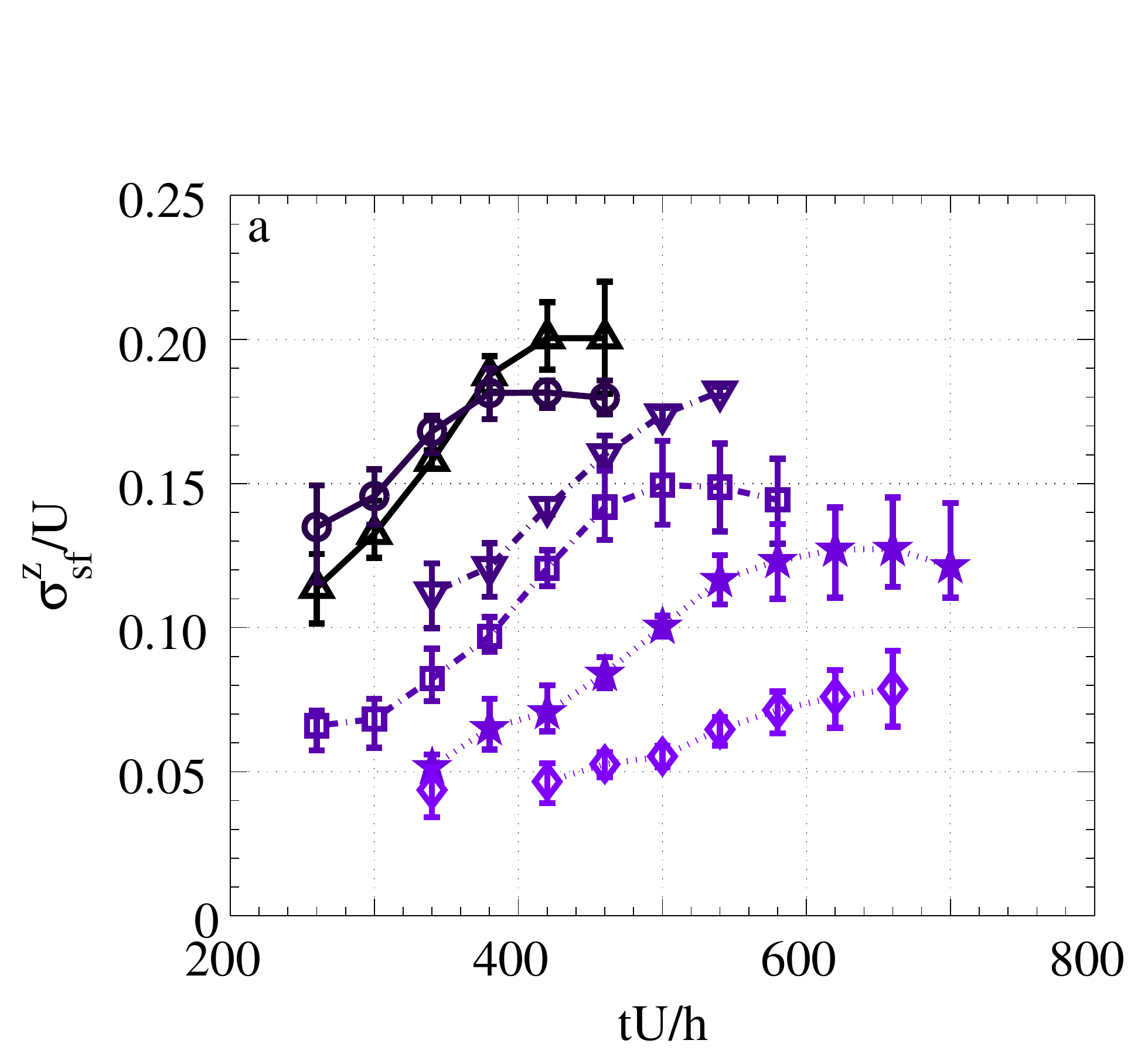}
	 \includegraphics[clip,trim = 4mm 0mm 0mm 30mm,width=0.495\linewidth]{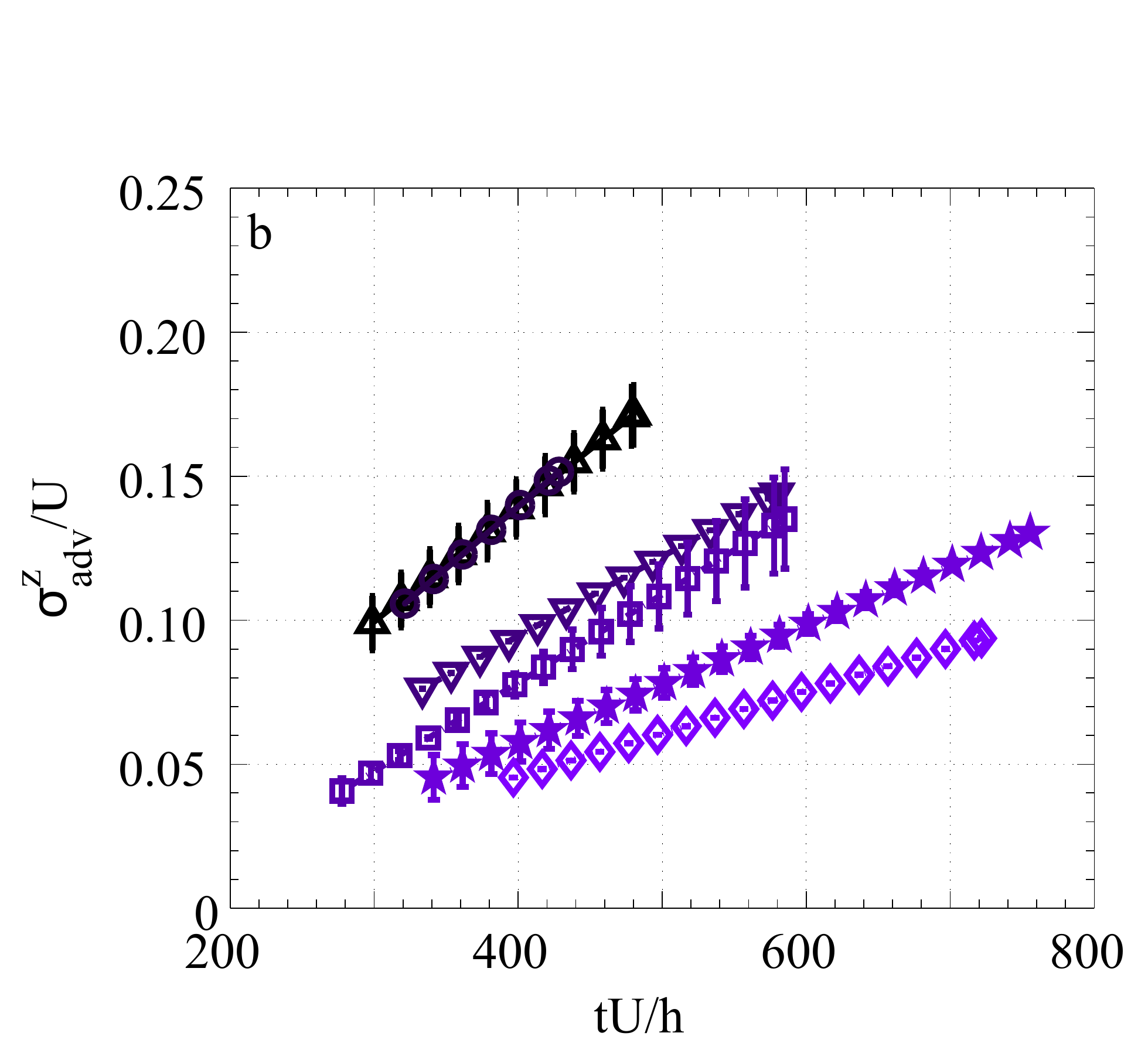}
	 \includegraphics[clip,trim = 4mm 0mm 0mm 30mm,width=0.495\linewidth]{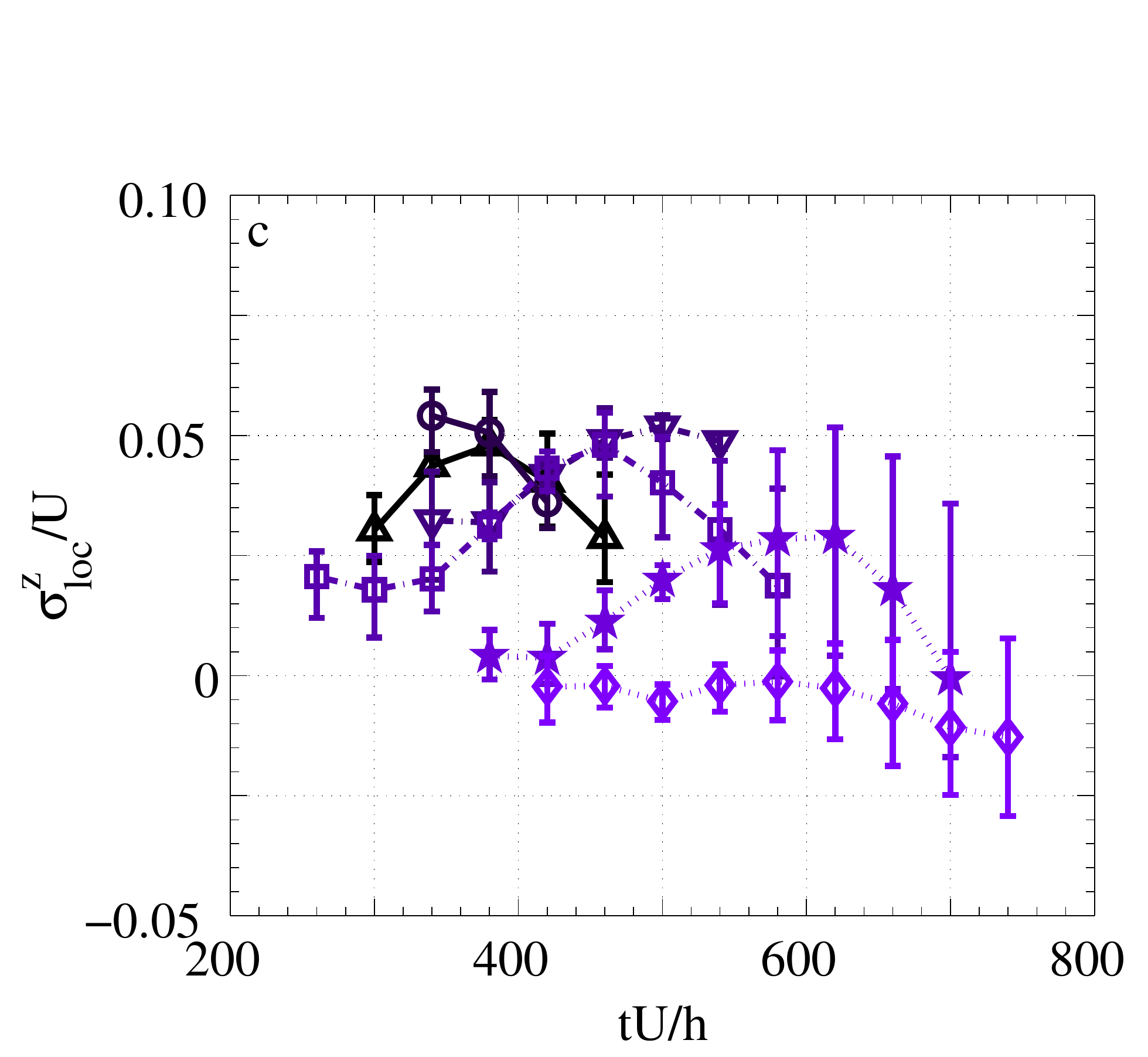}
	 \includegraphics[clip,trim = 4mm 0mm 0mm 30mm,width=0.495\linewidth]{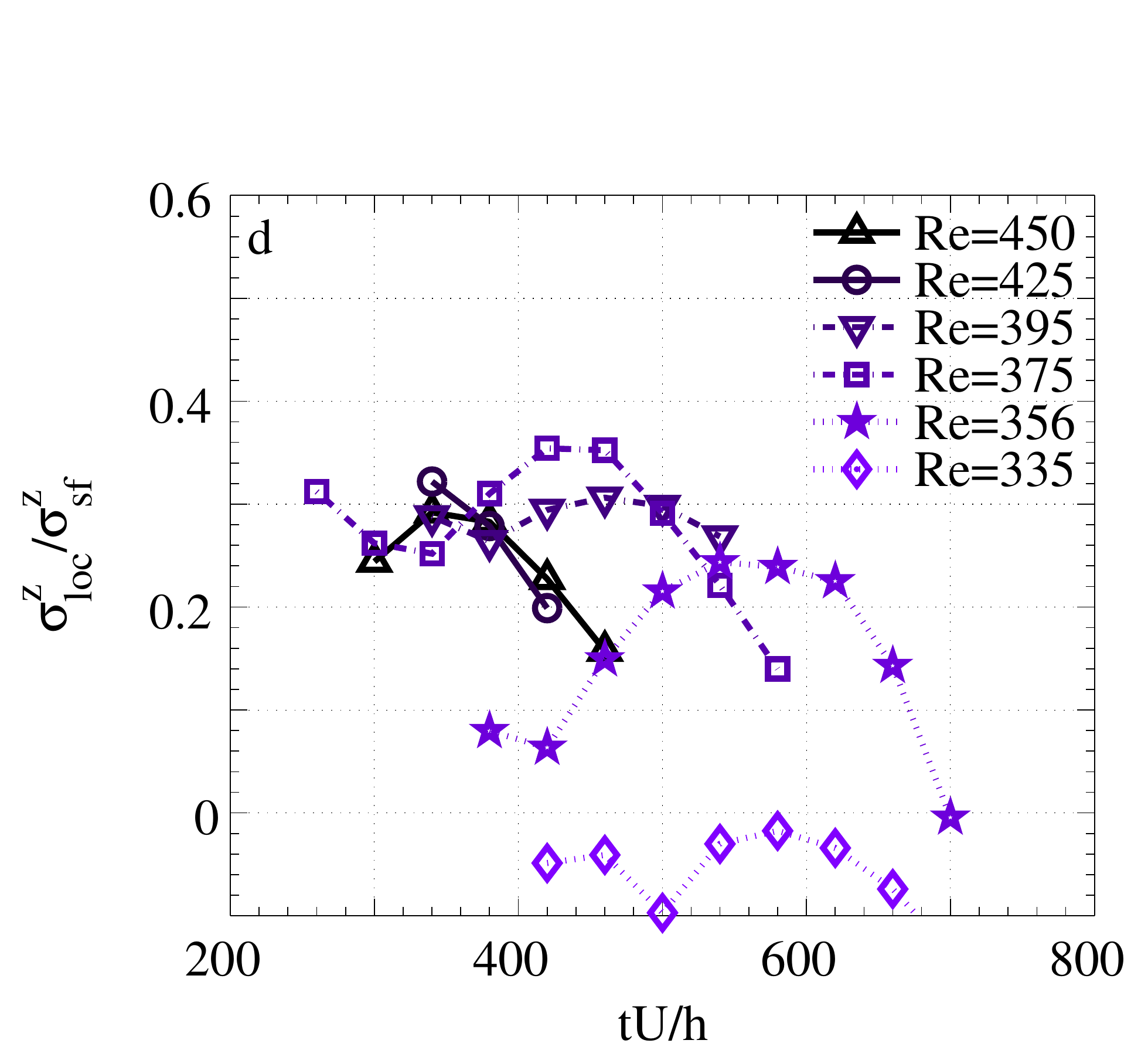}
		     \caption{Total spot growth rate $\sigma^z_{sf}$ (a), edge vortices advection velocity / advective growth rate $\sigma^z_{adv}$ (b), local growth rate $\sigma^z_{loc}$ (c) and ratio $\sigma^z_{loc}/\sigma^z_{sf}$ (d) as a function of time for various $Re$. Experimental data. Color online. \label{fig:sigma:expe}}
 	\end{center}
	\end{figure}

	Figure \ref{fig:sigma:expe} illustrates the time evolution of the four quantities defined in table~\ref{tab:sigma} for $6$ values of the Reynolds number spanning the range of $Re$ studied here. We do not present the $11$ available $Re$ values for the sake of clarity but the limiting cases detailed below take them into account.
	
	In figure \ref{fig:sigma:expe}(a), $\sigma _{sf}^z$ is seen to monotonically increase with time, while saturating at the end of the growth, resulting in $S$-shaped curves whose typical amplitude is $0.13$~$U$. When varying the Reynolds number, the overall shape of these curves does not change, but their starting point is higher and occurs at earlier time for larger $Re$ so that they can be deduced one from another by simple translations. The spot growth is faster at larger $Re$ and, with the type of perturbation used here, it also starts earlier, even in dimensionless units. This is probably due to faster development of the perturbation around the bead in the first stages of the experiment (see section \ref{sec:expe_prot}).
	
	In figure \ref{fig:sigma:expe}(b), $\sigma _{adv}^z$, the velocity of the vortices is seen to linearly increase with time. The rate of this increase grows with the Reynolds number so that, at the end of the growth process, vortices travel much faster at larger $Re$ while in the first steps of the growth dynamics, all vortices go roughly at the same speed ($0.05$~$U$ within our measurement accuracy), except at the very large $Re$ values. Note that the time span on which the velocity of the vortices is displayed is different from the displayed time span for the spot front position. This is because vortices are not always found in the space-time diagram at all times, either because they are not present, or because the visualization quality does not allow us to accurately track them. As $\sigma_{loc}^z$ and  $\sigma_{loc}^z/\sigma_{sf}^z$ are derived from $\sigma _{adv}^z$, the same remark obviously applies for them too.
	
	Due to the different dependencies for $\sigma_{sf}^z$ and $\sigma_{adv}^z$ on $Re$, their difference $\sigma _{loc}^z$ has more complicated dynamics that strongly depends on $Re$. The local contribution to the total growth is most of the time between $10\%$ and $30\%$ at any $Re$. This shows that most of the spot growth has to be attributed to the global mechanism as already proven from the time-average study by \cite{couliou15_POF}.	
	\subsection{Spreading in numerics}\label{sec:spread:num}
\begin{figure} 
	 \begin{center}
	 \includegraphics[clip,trim = 4mm 0mm 0mm 4mm,width=0.495\linewidth]{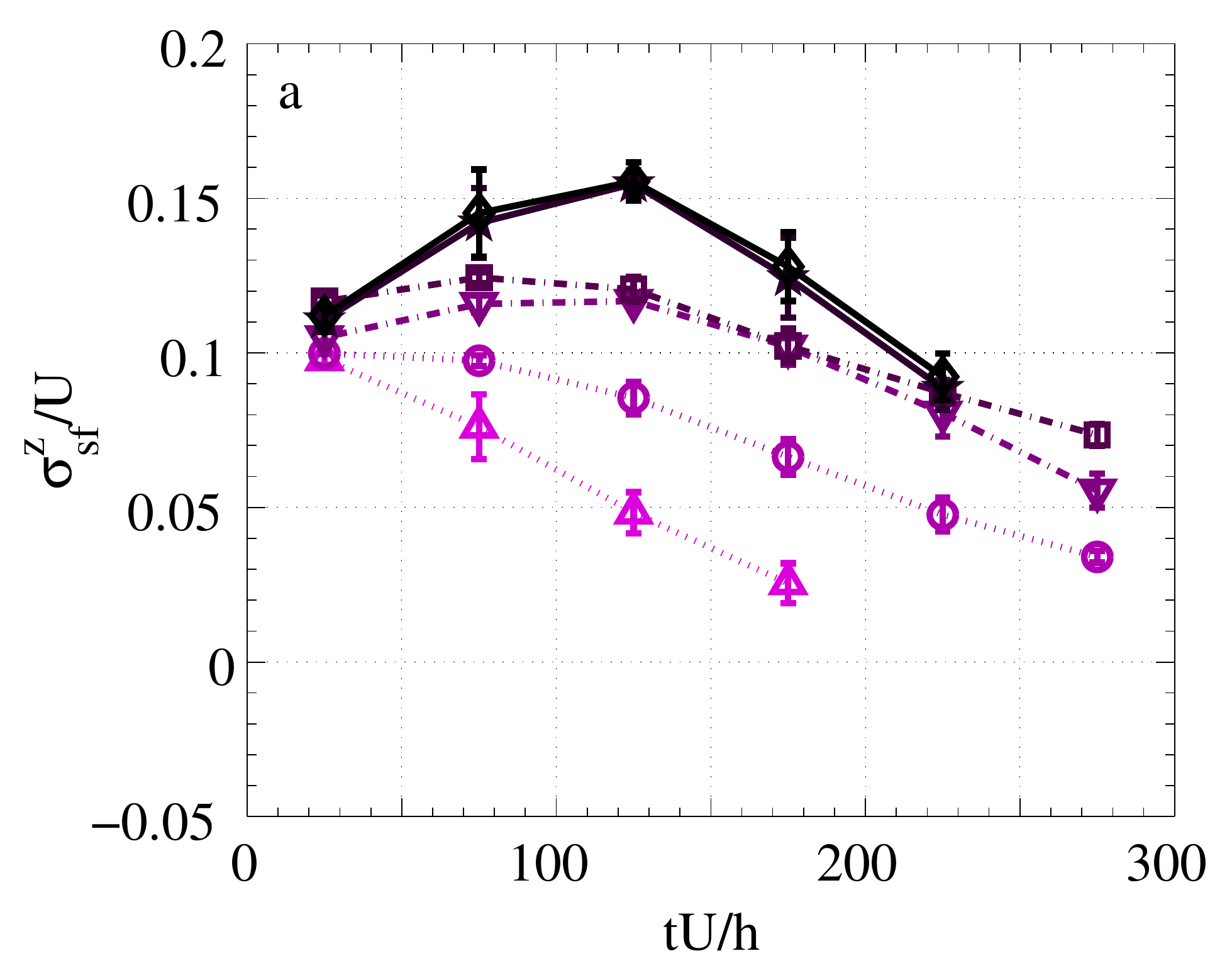}
	 \includegraphics[clip,trim = 4mm 0mm 0mm 30mm,width=0.495\linewidth]{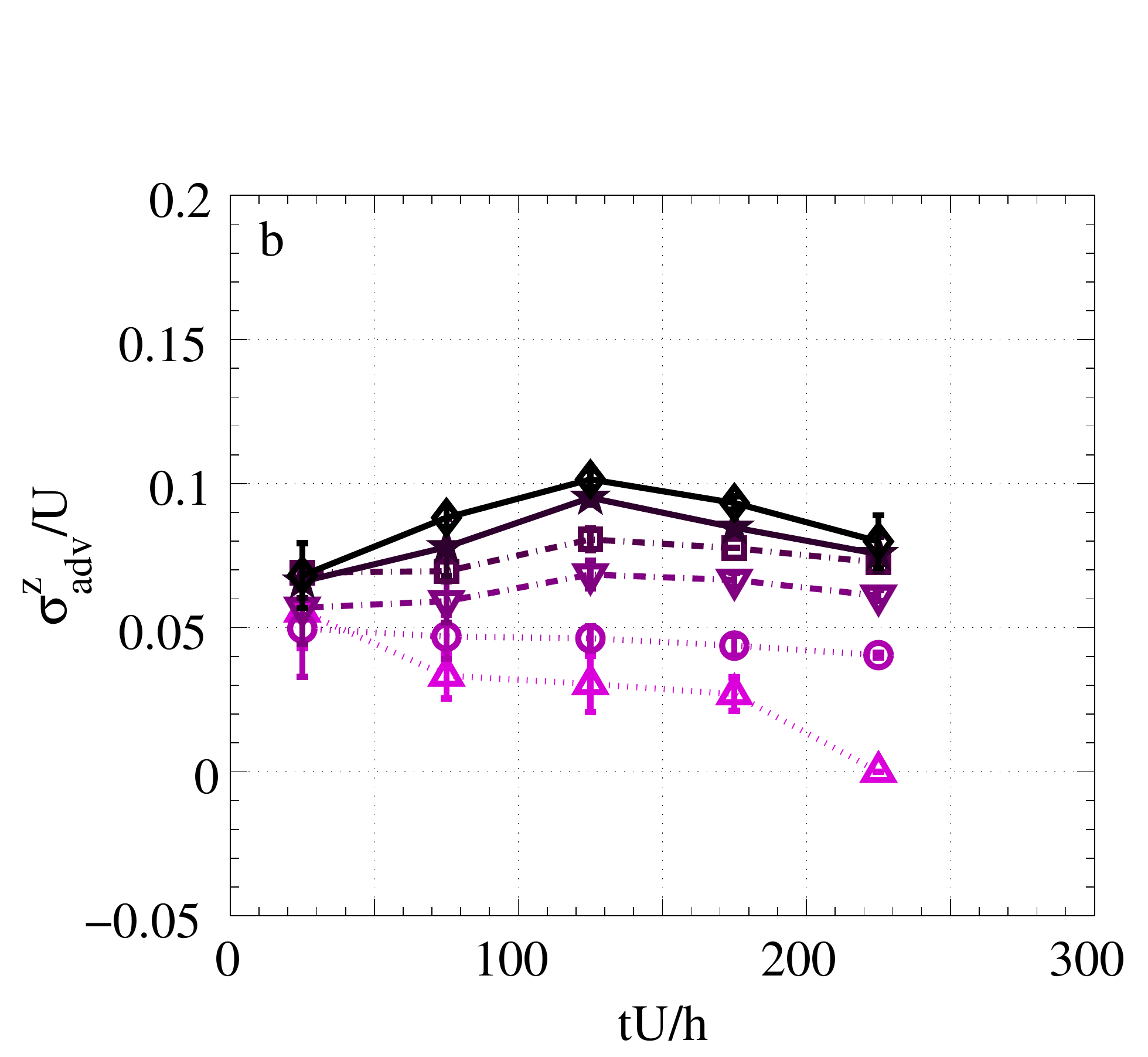}
	 \includegraphics[clip,trim = 4mm 0mm 0mm 4mm,width=0.495\linewidth]{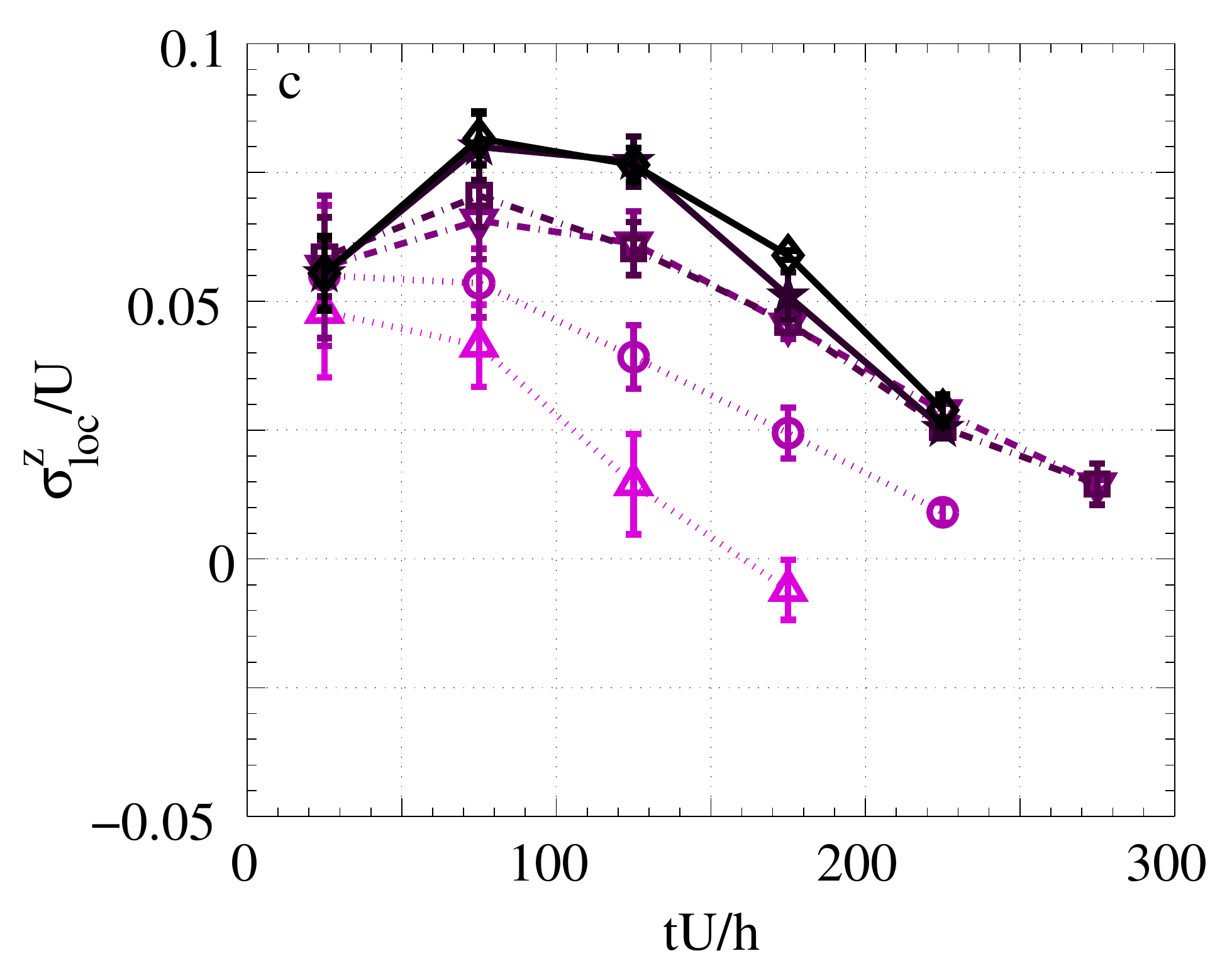}
	 \includegraphics[clip,trim = 4mm 0mm 0mm 4mm,width=0.495\linewidth]{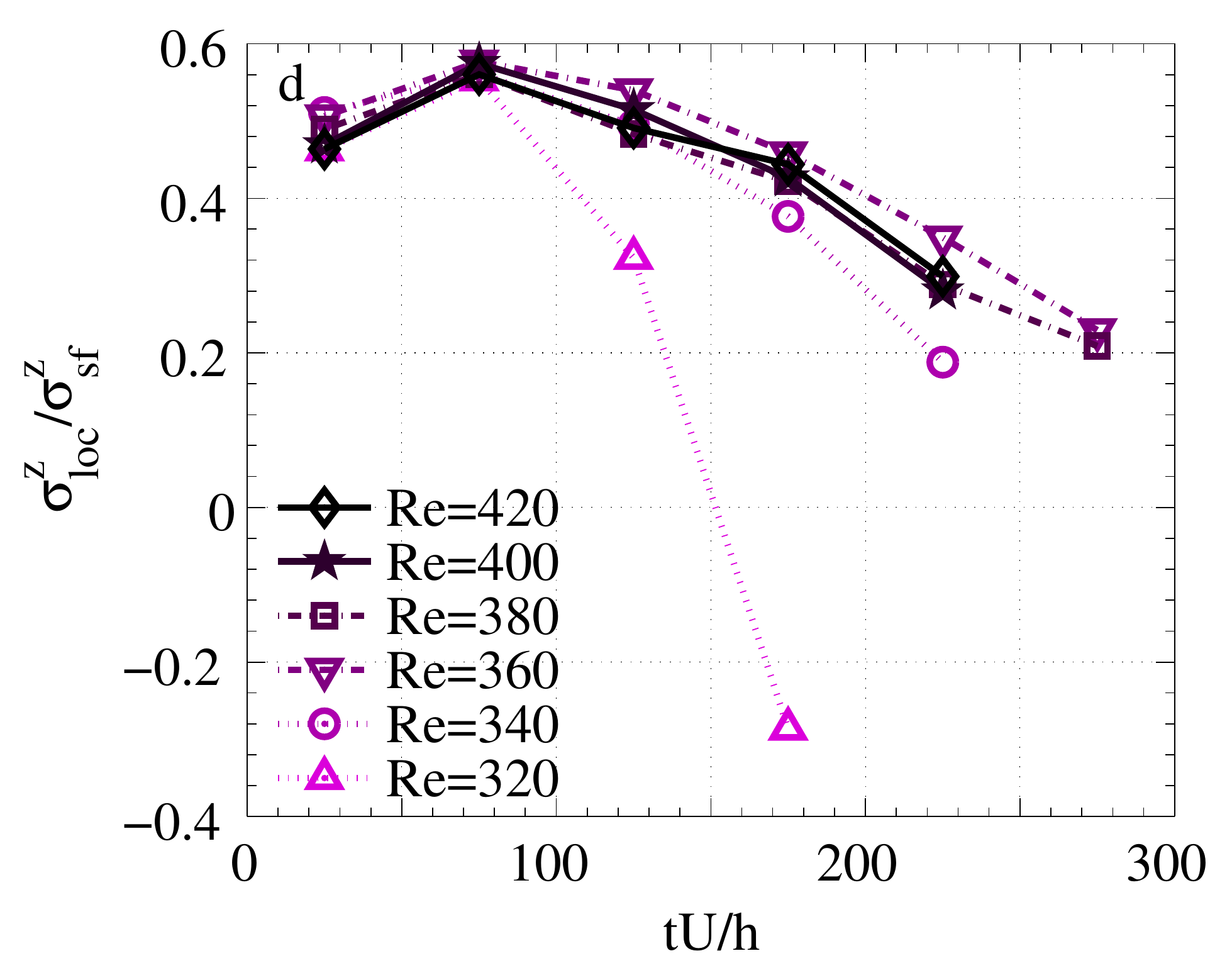}
	 \caption{Total spot growth rate $\sigma^z_{sf}$, edge vortices advection velocity / advective growth rate $\sigma^z_{adv}$ , local growth rate $\sigma^z_{loc}$ and ratio $\sigma^z_{loc}/\sigma^z_{sf}$ as a function of time for various $Re$. Numerical data. \label{fig:sigma:num}}
	 \end{center}
	\end{figure}

	Figure \ref{fig:sigma:num} gathers the time evolution of the three growth rates defined above and of their ratio for different Reynolds numbers in the numerical simulations. As a preliminary remark, we can note that all these figures differ from the experimental ones. Precise differences will be explained in this section. Note that for $Re=320$, spots grow during a first phase but eventually vanish. For $Re=340$,  spots arise in most cases but their sizes stay moderate and fluctuate.
	
	Contrary to what was observed in the experiment, the behaviour of $\sigma _{sf}^z$ strongly depends on $Re$. For $Re>360$, its maximum is reached at about $125$~$h/U$. Below this value, $\sigma _{sf}^z$ monotonically decreases with time and the decreasing rate is stronger with lower Reynolds number. Note that for $Re=320$, the spreading rate ends up negative since the considered spot actually vanishes (not presented). Regardless of the Reynolds number, the total growth rate $\sigma _{sf}^z$ is of the order of $0.12$~$U$ at short times and takes values between $0.05$~$U$ and $0.15$~$U$. This order of magnitude is similar to the experimental one.
	
	The quantity $\sigma^z_{adv}$ also displays a dynamics that is more complicated than in the experimental case and that is $Re$ dependent. It is roughly constant at low $Re$, slowly increasing at intermediate $Re$ values but presents a maximum for the larger $Re$. This maximum is once again reached around $125$~$h/U$. Regarding the order of magnitudes, they increase with $Re$ from $0.05$~$U$ to $0.1$~$U$. Once again, this order of magnitude is similar to the experimental one, although slightly smaller for larger $Re$.
	
	The quantity $\sigma_{loc}^z$ has time and $Re$ dependencies very similar to that of $\sigma_{sf}^z$, except that the maxima are reached at earlier times. As in the experimental case, its order of magnitude is around $0.05$~$U$.

	The trend of $\sigma_{loc}^z/\sigma_{sf}^z$  are almost identical regardless of the Reynolds number. They increase with time from values around $50\%$ to $60\%$ in the early stage of the growth before decreasing down to values around $20\%$. The order of magnitude is clearly smaller than in the experimental case.

	Coming back to the underlying mechanisms, these numerical results show that the global growth induced by large-scale flows dominates the growth rate at long times while local growth is much more active at the beginning of the growth process, leading to a balance between the two mechanisms. It is interesting to note that, when comparing the order of magnitudes seen in the experiment and in the simulation, they are very similar for the local growth ($\sigma_{loc}^z$) but clearly larger in the experiment for the global growth ($\sigma_{adv}^z$). This will be further discussed in section \ref{sec:disc}.

		\subsection{Large-scale flows}\label{sec:LSF}
	As explained in section \ref{sec:method}, large-scale flows and spot growth are not measured experimentally during the same experiment. Thus a straightforward temporal comparison of the associated dynamics is not possible. We have chosen to time-shift the large-scale flow quantities so that time $t=0$ corresponds to early stage of the growth presented above, at around one fourth of the overall process.
		\begin{figure}
	 \begin{center}
	 \includegraphics[width=0.495\linewidth]{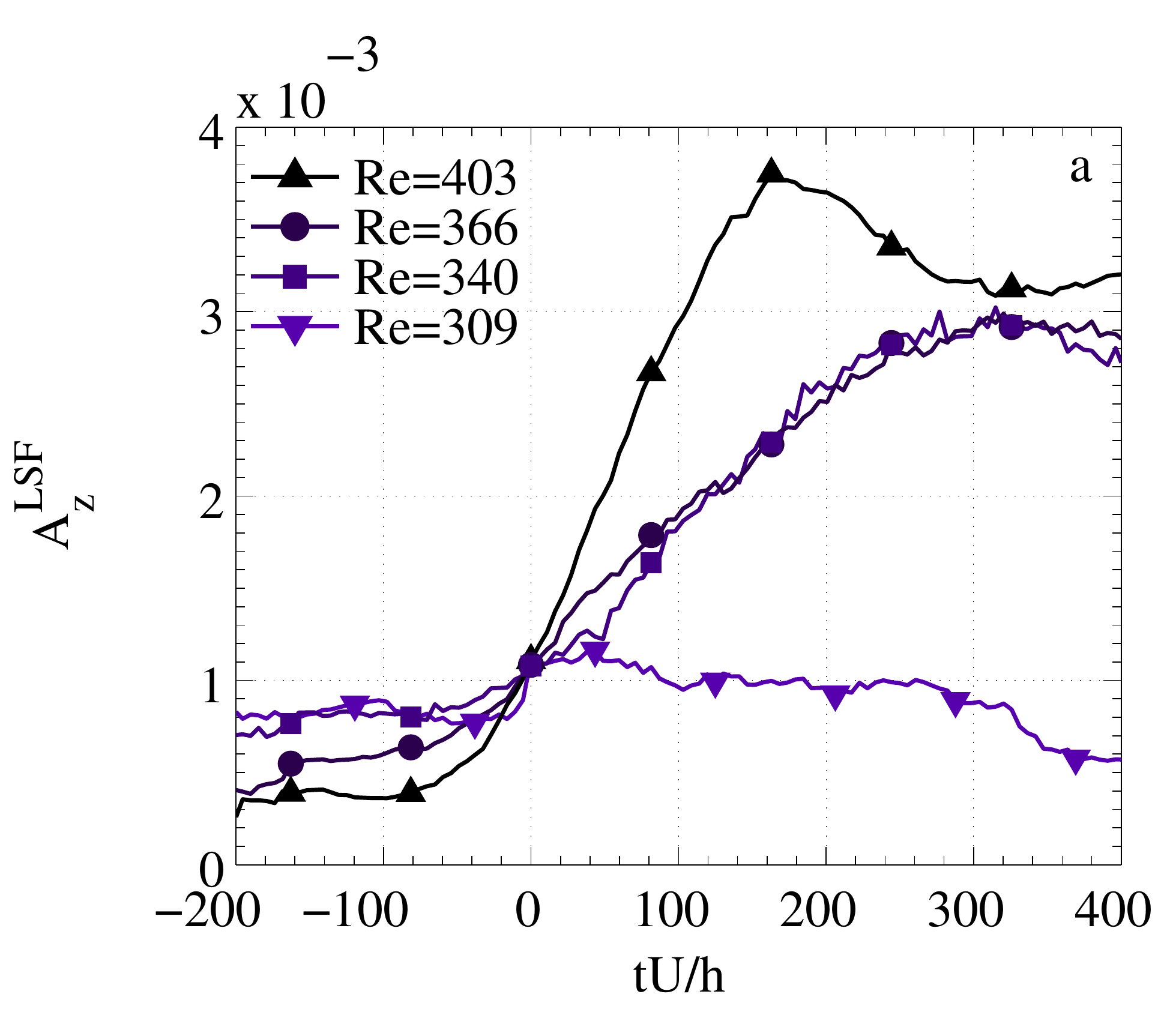}
	 \includegraphics[width=0.495\linewidth]{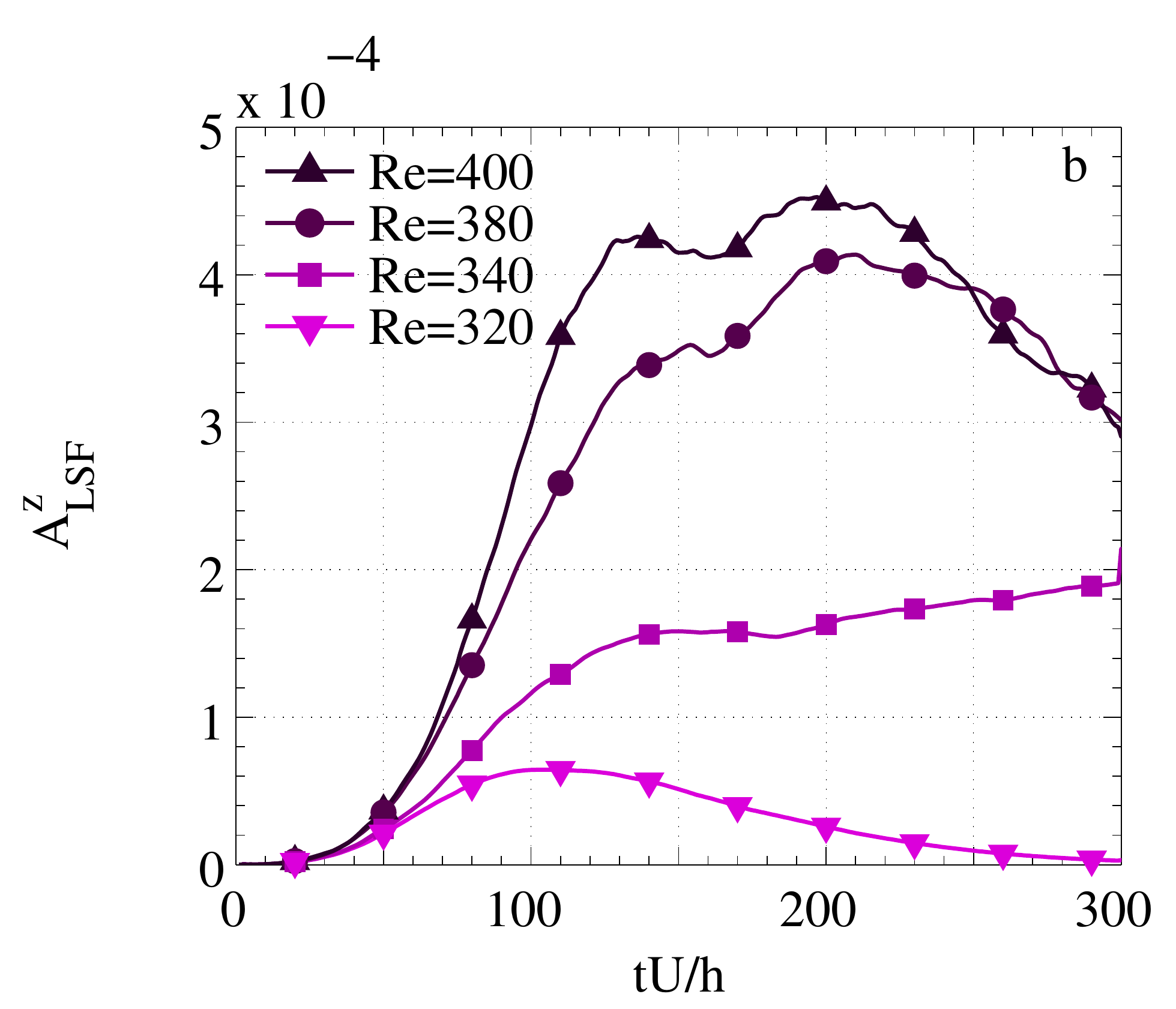}
	 \includegraphics[width=0.495\linewidth]{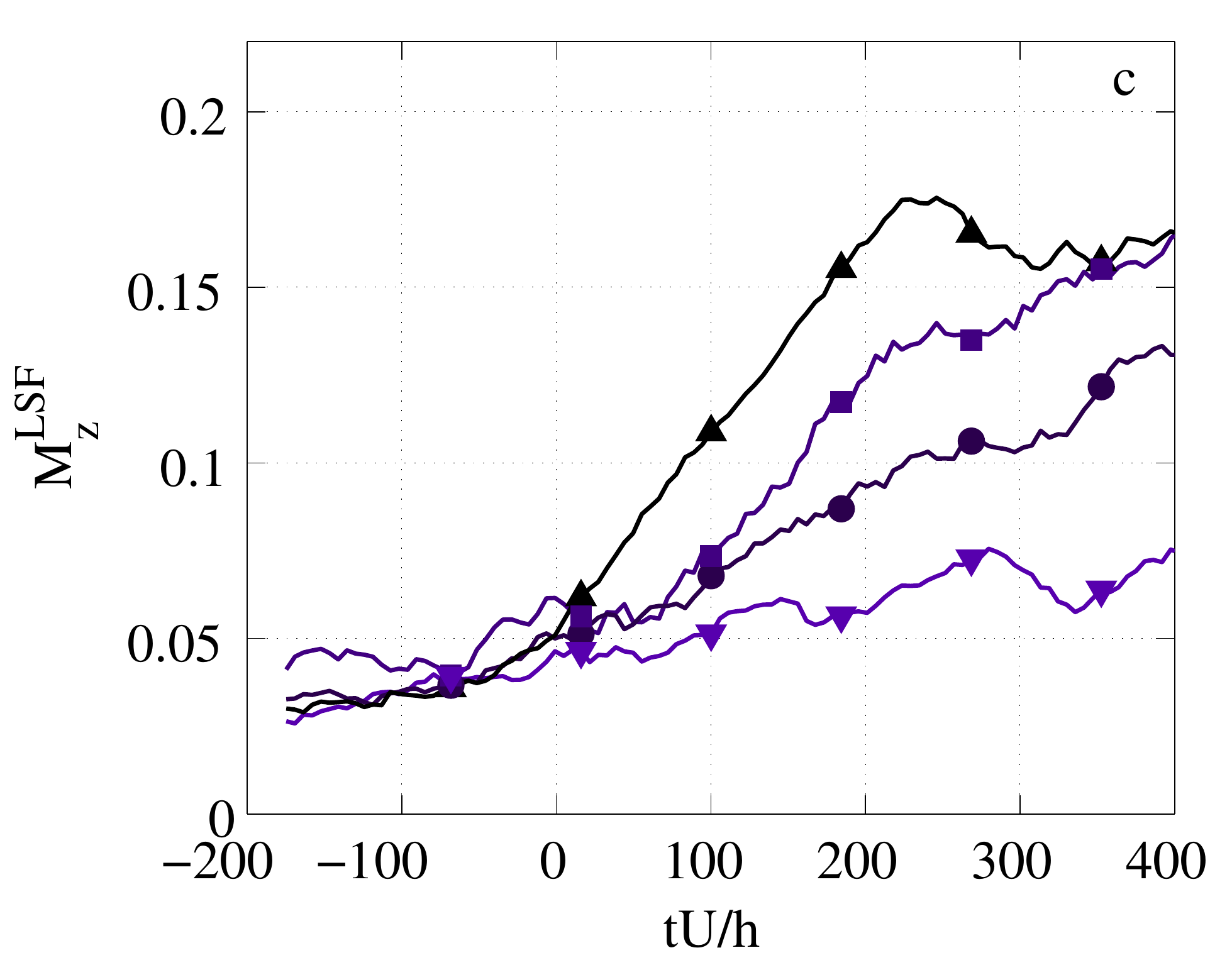}
	 \includegraphics[width=0.495\linewidth]{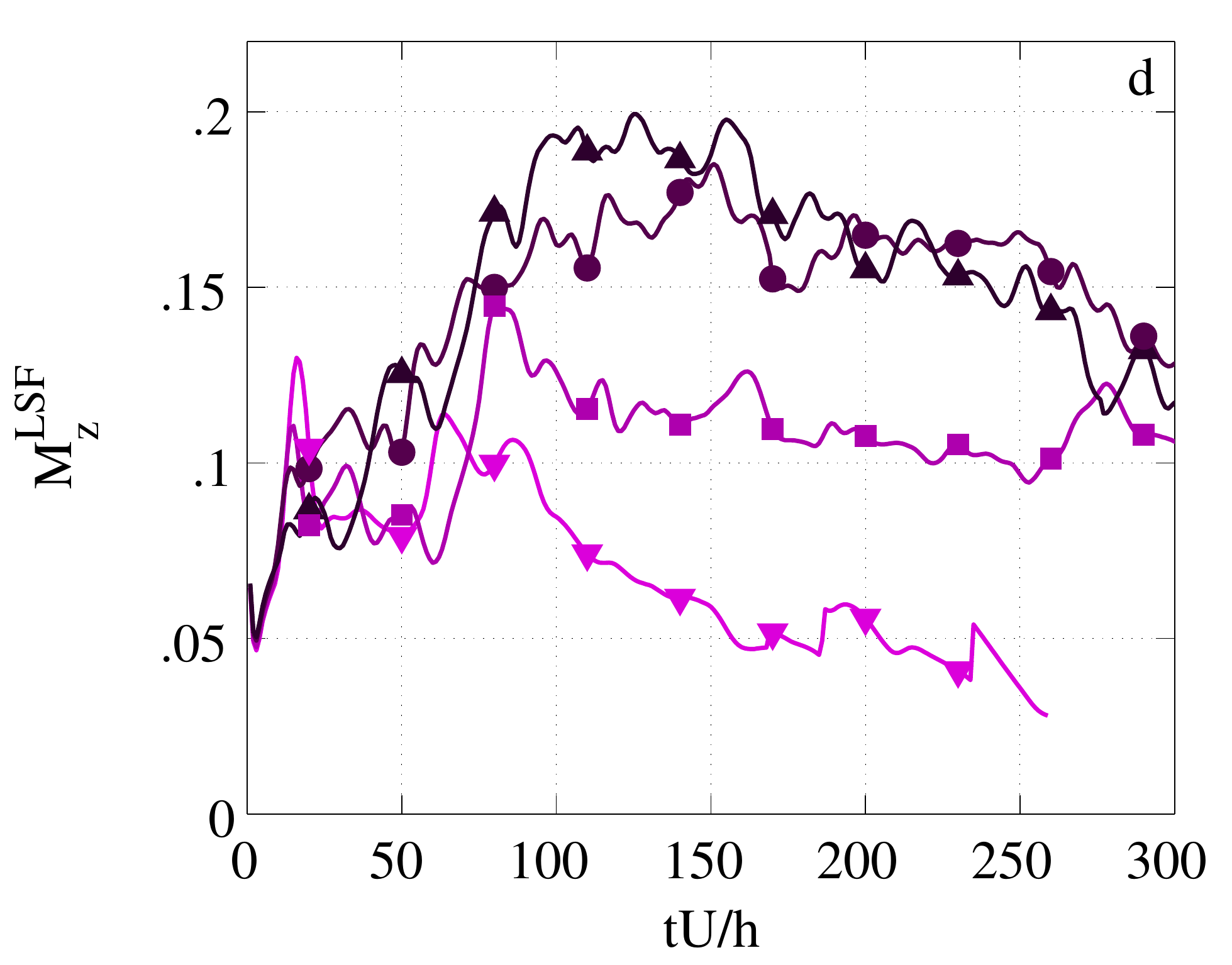}
	 \caption{$A^{LSF}_{z}$(top line) and $M_z^{LSF}$ (bottom line) as a function of time for different $Re$ in the experiment (left) and in the DNS (right). See text for precise definitions.\label{fig:maxUzLSF_exp}}
	 \end{center}
	\end{figure}
	
	In section \ref{sec:meth:lsf}, we have introduced two quantities associated with large-scale flows: $A_z^{LSF}$ and $M_z^{LSF}$. The quantity $A_z^{LSF}$ gives a global measure of the large-scale flow amplitude while the quantity $M_z^{LSF}$ is a local direct estimation of the order of magnitude of the advection of edge vortices by large-scale flows (see section \ref{sec:intro}). Figures \ref{fig:maxUzLSF_exp}(a) and \ref{fig:maxUzLSF_exp}(c) illustrate the evolution of these global and local measurements in our experiment on time scales relevant to the growth process as explained above. Both have relatively similar time dependencies. At low $Re$, they are fluctuating around a low constant value during the whole growth process. When $Re>330$, they increase from the beginning of the growth process and then saturate or slowly decay (actually, at later non-represented times, they all decay). For higher values of $Re$, the increase is faster and the decay starts sooner. Similar observations can be made for the numerical data presented in figure \ref{fig:maxUzLSF_exp}(b) and \ref{fig:maxUzLSF_exp}(d). The maximum of $max(U_z^{LSF})$ is reached at about $t\simeq 120-150$~$h/U$ when the spot front velocity was found to be maximal too.

	%
	%
	%
	%
	%
	%
\section{Discussion}\label{sec:disc}
	\subsection{Comparison to former work}
	The spot front velocity as a function of time for different $Re$ is represented in figure~\ref{fig:sigma:expe}. Contrary to some of the previous works that found the spot front velocity to be independent of time, particularly those by \cite{tillmark95_EPL} and by \cite{dauchot95a_POF}, these plots show a time dependent spot front velocity, even if our experimental setup is very similar to the ones cited. In fact, these researchers have followed spot growth over shorter times and have performed linear fits of the measured spot fronts equivalent to that presented here in figure \ref{fig:res:scatt}(a). Restricting the spot growth presented here to their first part, performing the same linear fitting seems reasonable since the non-linear evolution clearly appears only when considering the full temporal signal.
	
	On average, the spot front velocity $\overline{\sigma_{sf}^z}$ is found both in our experiment and in our simulations to be of the same order of magnitude than the one presented in \cite{dauchot95a_POF} even though slightly larger, particularly in the experimental case. The dependence of $\overline{\sigma_{sf}^z}$ on $Re$ is nevertheless similar in all cases as illustrated with the three parallel curves presented in figure \ref{fig:cmp_kth_saclay}.
	
	The velocities of edge vortices measured here as a function of both time and $Re$ have an order of magnitude similar to those found by \cite{dauchot95a_POF} and by \cite{tillmark95_EPL}. Unfortunately, the dependencies of these velocities with time or Reynolds numbers were not provided, so that the comparisons can not be pushed further. As we are, to our knowledge, the first to investigate the time evolution of the spot development in plane Couette flow and the link to with large-scale flow development, we cannot compare this part of our study to former contributions. The next paragraph is dedicated to a discussion of these aspects.
	\subsection{Experiments \textit{vs} numerics}
\begin{figure}
	 \begin{center}
	\includegraphics[clip,trim=0mm 10mm 0mm 0mm,width=0.8\textwidth]{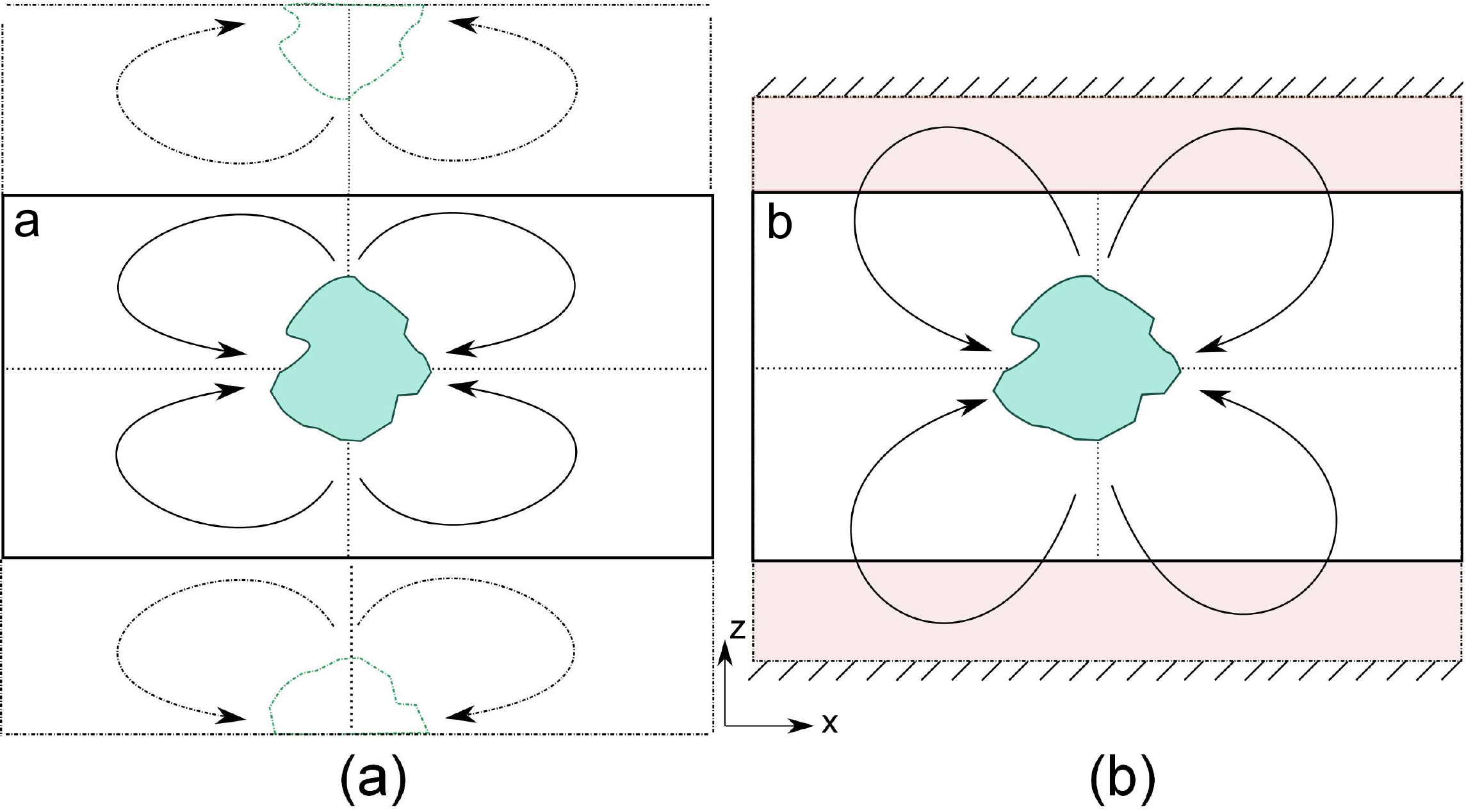}
	\caption{Schematic representation of the periodic boundary conditions in numerical simulation (a) and of water at rest and wall boundary conditions in the experiment (b).}
	\label{fig:why}
	 \end{center}
	
	\end{figure}

	As already mentioned in the last section, there are differences between the numerical and experimental velocity evolution presented here. When the numerical domain size is modified, we have seen in section \ref{sec:method} that the growth of turbulent spots is affected. When the spot fronts reach $0.87$~$L_z$ which corresponds to a distance to the domain boundary of $0.13$~$L_z$, the fronts slow down. We have thus restricted our analysis to regions far enough from the domain boundaries in order to limit the influence of this finite-size effect. We have done so for both numerical and experimental data as already explained. Nevertheless, boundary conditions are very different in both cases as sketched in figure~\ref{fig:why}. In simulations, they are periodic while in the experiment, they consist of walls where the spanwise velocity vanishes. The distance between these walls and the belt edges is about $L_z/7$. A buffer area thus exists between the shear zone and these walls. As a consequence, numerical and experimental fronts are analysed in physical zones of similar sizes but enclosed within very different boundary conditions. The main consequence of these differences is that large-scale flows associated with a numerical spot are tempered by their periodic images at times when, in the experiment, the buffer zone still allows these structures to develop. We indeed observe experimentally that the large-scale recirculation is actually present in the buffer zone as sketched in figure~\ref{fig:why}. Large-scale flows and thus the spot fronts are slowed down earlier in the numerical simulations than in the experiment. In the experiment, this slowing down should also be observed, but at later times. Since our analysis is restricted to the same physical space in both cases, and since the growth velocities have similar magnitudes, the numerical analysis spans a larger part of the whole dynamics than the experimental one. Indeed, we observe in figure~\ref{fig:sigma:num}(a) and \ref{fig:sigma:num}(b) that $\sigma^z_{sf}$ and $\sigma^z_{adv}$ decrease in the numerical case when they are still increasing in the experimental plots of figure~\ref{fig:sigma:expe}(a) and \ref{fig:sigma:expe}(b). Suppressing the buffer area in the experiment should lead to a more similar behaviour.
	%
	\subsection{Temporal dynamics}
	Considering numerical or experimental data, it is clear that at any time, the local and the global growth mechanisms are at work since $\sigma_{adv}^z$ and $\sigma_{loc}^z$ are non vanishing. Their relative contributions depend mostly on time. The quantities $\sigma_{adv}^z$ and $M^z_{LSF}$ have very similar temporal dynamics: in the experimental case, they are both increasing functions of time and they take values of the same order of magnitude in the range $0.05$-$0.17$~$U$. In the numerical case they exhibit maxima realized at the same time, around $t=125$~$h/U$. These strong similarities strengthen our conviction that the large-scale flows are responsible for the advection of the edge vortices and of the turbulent spot. From a more quantitative point of view, it is not easy to find a relevant measure of the large-scale flow intensity to be compared to the advection rate $\sigma_{adv}^z$. We have used the maximum, which is local and by nature an upper bound. A more refined estimation could be done by an integral measure (over $x$) at a given distance upstream the front and over a given streamwise length. This would imply the introduction of two free parameters and could lead to questionable results. Note nevertheless that the large-scale flow maximum is found to strictly remain at a distance $2-5$~$h$ upstream of the spot front at any time and for any $Re$ (figure not shown).
	\subsection{Growth mechanisms}
	Both in our previous work \cite{couliou16_PRE} and in the present work we have reported the existence of a global growth mechanism due to advection of the spot by the large-scale flows induced by the laminar-turbulent coexistence. This mechanism is associated with a growth rate $\sigma_{adv}^z$ that takes non-zero values at any $Re$ and at any time. This implies that large-scale flows are present throughout the growth process as already shown by direct PIV measurements of these flows by \cite{couliou15_POF}. As a consequence, the laminar profile outside of the turbulent spot differs from the linear Couette profile at any time and at any $Re$. A classical hypothesis to explain the growth of turbulent spots is, as detailed in the introduction, a local destabilization of the laminar profile at the spot spanwise tips. With our notation, the corresponding growth rate is $\sigma_{loc}^z$. At $Re=335$, $\sigma_{loc}^z$ equals zero at any time but large-scale flows are present. The configuration is such that the laminar profile is modified but does not trigger local growth. At $Re=346$, the local growth is active as $\sigma_{loc}^z$ is non-zero. We can thus infer the existence of a critical $Re$ value between $335$ and $346$ below which the laminar profile modification induced by the large-scale flow is not strong enough to trigger the instability. From the numerical simulations, the modified laminar flow is entirely known and can be extracted to perform linear stability analysis as was done by \cite{henningson87_JFM} for the plane Poiseuille case.
	
	As presented in the introduction, \cite{duguet11_PRE} suggest a local stochastic growth of the turbulent spot. This suggestion comes from observations in narrow domains where associated large scale flows do not develop. This stochastic growth is associated with a growth rate $\sigma_{stoch}^z$, whose order of magnitude is around $0.01$~$U$ in their narrow domains. This is one order of magnitude smaller than the two growth rates introduced in the present work. Actually,  \cite{duguet11_PRE} observe numerous streak-retreat events at any $Re$ while, in our numerical simulations, these events are very rare ($5$ streak-retreats over the $35$ simulations performed). If this stochastic growth exists in large domains, it can not explain the measured local growth rate that, as shown here, contributes to about $50\%$ of the total growth rate.
	\section{Conclusions and perspectives}\label{sec:persp}
	Spot growth in the transitional regime of plane Couette flow has  previously been studied numerically and experimentally by several researchers, revealing the existence of waves at the spot spanwise tips and pointing out the possible role played by large-scale flows surrounding the developing spot. The actual mechanisms at work to achieve the observed growth when $Re$ is large enough are still unclear although most of the former studies suggest a destabilization taking place at the spot spanwise tips. This suggestion is based on results time-averaged over the whole growth process. In our recently published work \citep{couliou16_PRE}, we have revealed the occurrence of streak nucleation within the turbulent phase pointing to a new global growth mechanism involving large-scale flows. In the present work, we have presented a temporal version of this study showing that the two suggested mechanisms are actually both present throughout the growth dynamics. At all times they contribute significantly to the total growth. Temporal comparison of the global growth rate $\sigma_{adv}^z$ and of measures of the large-scale flow intensities strengthen our interpretation that the large-scale flows are responsible for the advection of edge vortices and thus for the global spot growth rate. The role of nonlinear advection in the growth of turbulent spots was also noted from a recent model of pipe and channel flows proposed by \cite{barkley15_NAT}. From visualisations of both our experiments and simulations, we clearly see that large-scale flows change in shape and intensity during the reorganisation occurring when the spot is large enough to form the laminar-turbulent stripe pattern expected at long times. We suggest they also do play an important role in this reorganisation that we have planned to study. 
	
Regarding the spot growth in the streamwise direction, if the same kind of mechanisms act in this direction, due to their geometry, large-scale flows should be acting against the growth in this case. We are currently investigating these aspects from the data presented here. As plane Couette flow shares most of its features with the other extended shear flows, including the existence of large-scale flows around turbulent patches, it is worth performing the same analysis in plane Poiseuille and Taylor-Couette flows to see if the same conclusions can be formulated. 

\section*{Acknowledgments}
This project was supported by ANR Jeunes chercheurs/Jeunes chercheuses QANCOUET.
\bibliographystyle{jfm}
\bibliography{biblioJFM}


\end{document}